%% file: preprint.tex
\documentclass[numbers]{assets/matterlab}

\input{assets/shared-preamble.tex}
\setcitestyle{super,sort&compress}
\bibpunct{}{}{,}{s}{}{,}

\title{El Agente S\'olido: A New Age(nt) for Solid State Simulations}

\author[1, \orcidlink{0009-0006-8935-2302}]{Sai Govind Hari Kumar}
\author[1,2, \orcidlink{0009-0007-9131-4468}]{Yunheng Zou}
\author[1,6, \dagger, \orcidlink{0000-0003-3647-5000}]{Andrew Wang}
\author[1\dagger]{Jes\'us Vald\'es-Hern\'andez}
\author[1,6, \orcidlink{0000-0002-0802-9559}]{Tsz Wai Ko}
\author[1]{Nathan Yue}
\author[1]{Olivia Leng}
\author[1]{Hanyong Xu}
\author[2,5]{Chris Crebolder}
\author[*,1,2,3,4,5,6,7,8,9, \orcidlink{0000-0002-8277-4434}]{Al\'an Aspuru-Guzik}
\author[*,2,7, \orcidlink{0000-0002-8446-7956}]{Varinia Bernales}

\affiliation[1]{\addressCHEM}
\affiliation[2]{\addressCS}
\affiliation[3]{\addressMSE}
\affiliation[4]{\addressCHEMENG}
\affiliation[5]{\addressAC}
\affiliation[6]{\addressVECTOR}
\affiliation[7]{\addressCIFAR}
\affiliation[8]{\addressNVIDIA}
\affiliation[9]{\addressMS}

\contribution[\dagger]{Equal contribution}

\abstract{
\input{includes/include-abstract}
}

\date{\today}
\correspondence{\email{alan@aspuru.com} and \email{varinia@bernales.org}}

\begin{document}

\maketitle



\input{includes/include-body}


\section*{Acknowledgments}
\input{includes/include-acknowledgement}

\clearpage


{
\small
\bibliographystyle{unsrt}
\nocite{*}
\bibliography{references}
}


\clearpage

\appendix


\include{includes/include-appendix}


\end{document}

%% file: assets/shared-preamble.tex
\usepackage{microtype}
\usepackage{graphicx}
\usepackage{booktabs}
\usepackage{float}
\usepackage{xurl}
\usepackage{hyperref}
\usepackage{titlesec}
\usepackage{tabularx}
\usepackage{chemformula}
\usepackage{siunitx}
\usepackage{threeparttable}
\usepackage{xltabular}
\usepackage{ragged2e}
\usepackage{makecell}   
\usepackage{adjustbox}  
\usepackage{placeins}    

\usepackage{amsmath}
\usepackage{amssymb}
\usepackage{mathtools}
\usepackage{amsthm}
\usepackage{bm}

\usepackage[noabbrev,nameinlink]{cleveref}

\input{assets/math_commands}

\usepackage{orcidlink}
\usepackage{comment}
\usepackage[version=4]{mhchem}
\usepackage{longtable}
\usepackage{array}
\renewcommand{\arraystretch}{1.7}
\usepackage{multirow}           
\usepackage{amsfonts}           
\usepackage{nicefrac}
\usepackage{duckuments}         
\usepackage{thmtools,thm-restate}
\usepackage{enumitem}
\usepackage{xcolor,colortbl}
\usepackage{tikz}
\usepackage{tikz-cd}
\usepackage{caption}
\usepackage{subcaption}
\usepackage{listings}
\usepackage{gensymb}
\usepackage{textcomp} 
\usepackage[inkscapelatex=false]{svg}

\usepackage[most]{tcolorbox}
\newtcolorbox{noteBox}{colback=matterbg}

\newtcolorbox{hintBox}{colback=blue!10!white}

\newcommand{\prompt}[1]{\begin{noteBox}\underline{\textbf{Prompt}} #1 \end{noteBox}}

\newcommand{\elagenteQ}{\texttt{\textbf{El Agente Q}}}

\newcommand{\elagenteS}{\texttt{\textbf{El Agente S\'olido}}}
\newcommand{\solido}{\textbf{S\'olido}}
\newcommand{\estructural}{\texttt{\textbf{El Agente Estructural}}}
\newcommand{\QE}{\texttt{Quantum ESPRESSO}}
\newcommand{\qe}{\texttt{QE}}
\newcommand{\pormake}{\texttt{PORMAKE}}
\newcommand{\uma}{\texttt{UMA}}

\newcommand{\addressCHEM}{Department of Chemistry, University of Toronto,  80 St. George St., Toronto, ON M5S 3H6, Canada}
\newcommand{\addressAC}{Acceleration Consortium, 700 University Ave., Toronto, ON M7A 2S4, Canada}
\newcommand{\addressCS}{Department of Computer Science, University of Toronto, 40 St George St., Toronto, ON M5S 2E4, Canada}
\newcommand{\addressVECTOR}{Vector Institute for Artificial Intelligence, W1140-108 College St., Schwartz Reisman Innovation Campus, Toronto, ON M5G 0C6, Canada}
\newcommand{\addressMSE}{Department of Materials Science \& Engineering, University of Toronto, 184 College St., Toronto, ON M5S 3E4, Canada}
\newcommand{\addressCHEMENG}{Department of Chemical Engineering \& Applied Chemistry, University of Toronto, 200 College St., Toronto, ON M5S 3E5, Canada}

\newcommand{\addressCIFAR}{Canadian Institute for Advanced Research (CIFAR), 661 University Ave., Toronto,
ON M5G 1M1, Canada}
\newcommand{\addressNVIDIA}{NVIDIA, 431 King St W \#6th, Toronto, ON M5V 1K4, Canada}
\newcommand{\addressMS}{Institute of Medical Science, 1 King's College Circle, Medical Sciences Building, Room 2374, Toronto, ON M5S 1A8, Canada}

\newcommand{\acknowAC}{This research is part of the University of Toronto’s Acceleration Consortium, which receives funding from the CFREF-2022-00042 Canada First Research Excellence Fund. }
\newcommand{\acknowGEN}[1]{This research was enabled in part by support provided by #1 and the Digital Research Alliance of Canada (\url{alliancecan.ca}).}



%% file: assets/math_commands.tex
\usepackage{amsmath,amsfonts,bm}

\def\eqref#1{equation~\ref{#1}}

\def\1{\bm{1}}

\DeclareMathAlphabet{\mathsfit}{\encodingdefault}{\sfdefault}{m}{sl}
\SetMathAlphabet{\mathsfit}{bold}{\encodingdefault}{\sfdefault}{bx}{n}



%% file: includes/include-abstract.tex
Quantum chemistry calculations are a key component of the materials discovery process. The results from first-principles explorations enable the prediction of material properties prior to experimental validation. Despite their impact, the practical use of first-principles methods remains limited by the expertise required to design, execute, and troubleshoot complex computational workflows. Even when workflows are successfully built, they are sometimes rigid and not adaptable to different use cases. Recent advances in large language models (LLMs) and agentic systems offer a pathway to flexibly automate these processes and lower barriers to entry. Here, we introduce \elagenteS{}, a hierarchical multi-agent framework for automating solid-state quantum chemistry workflows using the open-source \QE{} simulation package. The framework translates high-level scientific objectives expressed in natural language into end-to-end computational pipelines that include structure generation, input file construction, workflow execution, and post-processing analysis. \elagenteS{} integrates density functional theory with phonon calculations and machine-learning interatomic potentials to enable efficient and physically consistent simulations. Extensive benchmarking and case studies demonstrate that \elagenteS{}  reliably executes a wide range of solid-state calculations, highlighting its potential to improve reproducibility and accelerate computational materials discovery.

%% file: includes/include-body.tex
\section{Introduction}

Materials discovery has been a key driver of technological progress and has been key to improving human lives since the dawn of the industrial age~\cite{ozin_is_2025}. The search for new materials continues to this day and remains important for advancing fields such as energy storage \cite{kirklin_high-throughput_2013, chen2024accelerating, mortazavi2025recent, kahle_high-throughput_2020, laskowski_identification_2023, angelis_energy-gnome_2025}, photovoltaics \cite{yuan_discovery_2024, choudhary2019accelerated}, and catalysis \cite{hari_kumar_computational_2024, ma2020machine, back_discovery_2020, sun_covalency_2020, tran_open_2023}, which are necessary to facilitate a transition to a sustainable world. Materials discovery has traditionally been facilitated by a mix of chemical intuition and serendipity \cite{ozin_is_2025}. This, however, is a slow, laborious process that is inefficient at exploring the chemical space. According to one estimate by Walsh and coworkers, it consists of more than 32 million ternary inorganic phase materials, even after constraining this space by imposing charge-neutrality and electronegativity rules \cite{davies_computational_2016}. It is possible to navigate a much larger fraction of this space using high-throughput computational screening to narrow down a potential list of novel candidates, which can then be synthesized and characterized in the lab \cite{merchant2023scaling, chen2024accelerating, tran_open_2023, Jain2013MaterialsProject}. Such an approach has already been used to discover new materials in fields such as catalysis \cite{hari_kumar_computational_2024,back_discovery_2020, sun_covalency_2020}, Li-ion batteries \cite{hautier_novel_2011, kirklin_high-throughput_2013, angelis_energy-gnome_2025, chen2024accelerating}, thermoelectrics \cite{zhu_computational_2015, xi_discovery_2018, pohls_experimental_2021}, organic light-emitting diodes \cite{gomez-bombarelli_design_2016, jorner_ultrafast_2024, an_machine_2026}, photovoltaics  \cite{yuan_discovery_2024, fabini_candidate_2019, choudhary2019accelerated}, and transparent conducting oxides \cite{yim_computational_2018, woods-robinson_designing_2023, youn_large-scale_2019}. Computational chemistry is, therefore, a powerful tool in any materials scientist's toolkit, capable of significantly accelerating the pace of materials discovery.

Despite the demonstrated value of computational density functional theory (DFT) in materials discovery, many experimental materials scientists still do not incorporate it into their workflows. This is largely because running first-principles electronic-structure simulations demands substantial computational and technical expertise. Widely used packages such as \QE{} \cite{Giannozzi2009QE}, \texttt{VASP} \cite{Kresse1996CMS}, \texttt{CRYSTAL} \cite{erba_crystal23_2023}, \texttt{ABINIT} \cite{Gonze2009ABINIT}, \texttt{CASTEP} \cite{Clark2005CASTEP},  \texttt{JDFTx} \cite{Sundararaman2017JDFTx}, among others, enable quantitative prediction of materials properties using DFT and related methods, but remain difficult to use for a significant portion of the community \cite{aspuru2025rise}. A major challenge is designing and executing simulations that are computationally well-defined and numerically stable. Users must correctly construct input files, master program-specific syntax, and operate in non-standard computing environments, typically UNIX, creating a steep learning curve for the non-expert. As a result, researchers frequently spend considerable time on setup and troubleshooting, diverting effort away from framing and investigating the underlying scientific questions.
Autonomous systems guided by large language models (LLMs) offer an alternative paradigm to address this limitation. Within this framework, expert agents are designed to interpret high-level scientific objectives expressed in natural language and to autonomously orchestrate the sequence of computational tasks required to achieve them \cite{zou2025agente, aspuru2025rise, duan2025boosting, perezsanchez2026elagentequntur, bai_agente_2026, choi2026elagenteestructural,gustin2026elagentecuantico}. These agents can generate input files, select appropriate computational parameters, execute simulations, and adapt workflows based on intermediate results, all while abstracting away low-level technical details from the user. By separating high-level scientific objectives from their computational implementation, LLM-driven agents enable researchers to focus primarily on the analysis, validation, and physical interpretation of results. Beyond improving efficiency, this paradigm has the potential to enhance reproducibility, lower barriers to entry in computational materials science, and facilitate the systematic exploration and proposal of novel materials with targeted properties.

Recently, there has been a rapid growth in the number of LLM-based agentic frameworks for materials science \cite{mofgen2025, MAPPS2025, dreams2025, SciLink2025, chatmof2024, catmaster2026, vaspilot2025, adsorbagent2025, AGAPIAgents2025, genius2025, crystalyse2025, AURA2025, LLMatDesign2024, matsciagent2025, VrizaEtAl2025, atomagents2024, Masagent2025, yang_quasar_2026, KDensePaper2025, KDenseCompany2025, ParamusCompany2025}. A summary of their main features, primary external code(s) used, and computational benchmarks can be found in Table \ref{tab:agents}. However, the consistency and reproducibility of results across multiple trials of prompting was rarely evaluated, and the ability to generate more exotic structures such as disordered multicomponent materials (i.e. battery electrodes) has yet to be demonstrated. For example, the most recent release, QUASAR \cite{yang_quasar_2026}, showcased the agent's ability to handle well 9 different exercises from simpler tasks to research level problems, but like in many other reports, the results were only from a single trial. Given that LLMs are fundamentally non-deterministic, we believe that demonstrating the consistency and reproducibility of results across multiple trials is an important aspect of evaluating the performance of agentic frameworks.  


 Here we introduce \elagenteS{}, a hierarchical multi-agent framework for autonomous first-principles materials modelling that integrates \QE{} (\qe{}) within a unified, language-driven workflow. We demonstrate \elagenteS{}'s ability to consistently and reproducibly perform 10 iterations of 7 benchmark exercises, as well as showcase its ability to handle more exotic structures like Li-ion battery cathodes as well as other complex tasks that have previously been demonstrated, such as constructing slabs with adsorbed intermediates and generating reticular frameworks such as metal-organic frameworks (MOFs) and covalent-organic frameworks (COFs).
 
\begin{figure}[htbp]
    \centering
    \includegraphics[width=0.75\linewidth]{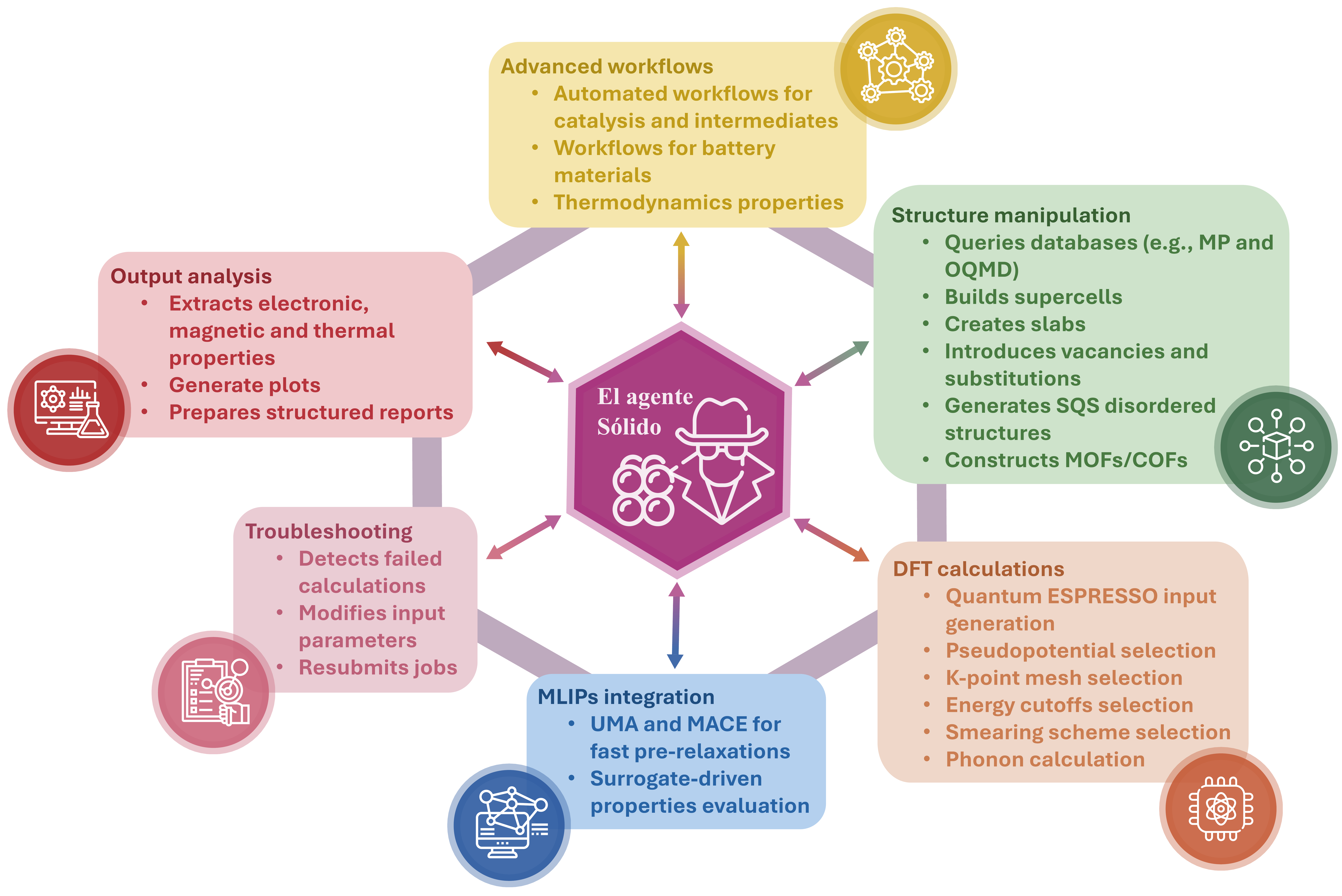}
    \caption{Core computational capabilities of \elagenteS{}, spanning structure generation, simulation, analysis, and advanced materials workflows.}
    \label{fig: capabilities}
\end{figure}

 \elagenteS{} can autonomously query external materials databases, generate and manipulate atomic structures (including vacancies, substitutions, supercells, slabs, disordered special quasirandom structures (SQS), and generation of porous materials), and select appropriate computational and physical parameters for each task. \elagenteS{} executes DFT workflows encompassing structural relaxations, self-consistent and phonon calculations, electronic band structures, densities of states, thermodynamic stability analyses, and surface adsorption studies. In addition, \elagenteS{} integrates state-of-the-art pretrained machine-learning interatomic potentials (MLIPs) with broad coverage across the periodic table~\cite{wood2025family, batatia2023foundation}, enabling large-scale calculations and significantly accelerating \textit{ab initio} simulations through MLIP-based pre-optimization, Figure \ref{fig: capabilities}. Across 7 benchmarking exercises, each repeated at least 10 times with the same prompt, \elagenteS{} achieved an average score of 97.9\% based on rubrics designed by computational chemists that graded its ability to set up complex workflows to answer the prompt and arrive at the correct answer. Additionally, we also devised four case studies demonstrating \elagenteS{}'s utility for materials discovery. Three out of the four case studies focus on materials discovery in three hot areas of materials research: lithium-ion batteries, catalysis and reticular frameworks like MOFs and COFs. In the final case study, we demonstrate \elagenteS{}'s integration with Phonopy by using it to calculate phonon bandstructures of materials as well as calculate thermal properties like heat capacity and thermal expansion coefficient. We demonstrate \elagenteS{}'s ability to construct adsorbed intermediates on surfaces to determine the theoretical overpotential of the oxygen evolution reaction, produce phonon band structures and compute thermal properties, construct SQS structures of disordered materials like NMC-811 and estimate their delithation voltage profile, and construct MOFs and COFs using \pormake{} and calculate their properties. Throughout these exercises we show that by combining autonomous decision-making with end-to-end workflow orchestration, \elagenteS{} enables systematic, reproducible, and scalable exploration of materials with targeted properties while substantially reducing manual intervention.

\section{Methodology Section}
\subsection{Architecture of \elagenteS{}}

\elagenteS{} shares borrows and extends the underlying cognitive architecture as \elagenteQ{}  \cite{zou2025agente}. The main difference between \solido{} and the original \elagenteQ{} is the composition of the procedural memory, which had to be modified in order to accomplish tasks relevant to solid-state chemistry, Figure \ref{fig: flowchart_architecture}(a). The procedural memory of \elagenteS{} consists of a hierarchical multi-agent team that cooperates to accomplish a computational chemistry task. Each agent, defined as a node within this procedural memory, is endowed with specialized context that defines the role of the agent, semantic memory that reinforces role adherence, and a set of callable modules that are either other agents below it in this hierarchical team that it's able to delegate tasks to, or specialized Python code that are used as tools which allow each subagent to accomplish very specific tasks, this is similar to the strategy used in \estructural{}\cite{choi2026elagenteestructural}.

At the top of this hierarchical team is the \texttt{Computational Chemist Agent}, who receives requests from the human scientist(s) and plans high-level workflows to fulfill them. It then delegates different aspects of the workflow to the following four subagents:
\begin{itemize}[leftmargin=2em]
    \item \texttt{Geometry Generator Subagent}: Responsible for generating the initial geometries of any material before any MLIP or DFT calculations are performed. This subagent can query databases such as OQMD~\cite{kirklin_open_2015} or the Materials Project~\cite{Jain2013MaterialsProject} for initial structures. It can create molecules from a SMILES string~\cite{noauthor_rdkit_nodate} or its IUPAC name~\cite{lowe_chemical_2011}. It can manipulate structures by creating supercells, vacancies, and substituted structures \cite{ong_python_2013}, and create disordered structures as special quasirandom structures (SQS)~\cite{angqvist_icet_2019, van_de_walle_efficient_2013, zunger_special_1990}. It also has access to collaborate with two subagents: (1) the \texttt{Surface Generator Subagent}, which can create slab structures with adsorbed molecules on the surface, and (2) the \texttt{MOF Generator Subagent}, which can create metal--organic frameworks.
    
    The \texttt{Surface Generator Subagent} is capable of generating a slab of the original structure when given a Miller index and a bulk structure~\cite{tran_surface_2016, sun_efficient_2013}. This subagent can also generate multiple slabs when given a maximum index and asked to generate all symmetrically distinct Miller indices. Additionally, it can adsorb small molecules on the surface of slabs to achieve the specified surface coverage. On the other hand, the \texttt{MOF Generator Subagent} can operate in two ways: either by querying them from the QMOF database~\cite{rosen_machine_2021} or by constructing them with \pormake{} \cite{lee_computational_2021}. If given a \texttt{MOFkey} or a \texttt{MOFid}~\cite{bucior_identification_2019}, the \texttt{MOF Generator Subagent} can query a specific structure from the QMOF database or construct it using the \pormake{} software, given a topology in which the secondary building unit and organic linker are represented by nodes and edges. This subagent will search \pormake{}'s database for nodes and edges compatible with the user's requested topology and generate multiple MOF structures.
    
    The initial structures generated by the \texttt{Geometry Generator Subagent} and its subagents can be pre-relaxed using reliable foundation MLIPs, including the Universal Model for Atoms (\uma{}) \cite{wood2025family} and message-passing atomic cluster expansions (MACE) \cite{batatia2023foundation} by the \texttt{Geometry Generator Subagent}, to accelerate subsequent DFT calculations.
    
    \item \texttt{DFT Subagent}: Responsible for running DFT calculations and also \texttt{Phonopy}~\cite{togo_implementation_2023, togo_first-principles_2023} calculations based on any geometries the \texttt{Geometry Generator Subagent} may create. It has access to three main subagents: i) The \texttt{Input File Generator Subagent}, which is primarily responsible for generating input files for \QE. This subagent has access to 13 other subagents to help it determine appropriate values for different input tags across the blocks of a typical \QE{} input file. ii) The \texttt{\qe{}-Running Subagent}, which runs the \QE{} calculation once the input file has been generated. This subagent sets appropriate parallelization parameters and verifies that essential files, such as pseudopotentials, are present and correct, ensuring a successful run. Once it has confirmed that all essential features of the calculation are present in the directory and selected optimal parallelization parameters, it runs a \QE{} calculation via \texttt{SLURM}. iii) the \texttt{Phonopy Running Subagent}, which primarily runs \texttt{Phonopy} calculations to generate the phonon band structure and calculate the material's thermal properties. Overall, the \texttt{DFT Subagent} is the key responsible for troubleshooting a calculation when it fails.
    \item \texttt{File I/O Subagent}: Primarily responsible for creating directories and subdirectories and organizing input and output files with a reasonable and human-readable naming convention into them.
    \item \texttt{Output Analyzer Subagent}: Responsible for analyzing the output of a \QE{} calculation. This subagent can parse output files to extract relevant information, such as optimized coordinates, final electronic energy, and other properties, from a calculation, and then report and plot band structures~\cite {hinuma_band_2017}, density of states, formation energy, and the bulk modulus.
\end{itemize}
\begin{figure}[htbp]
    \centering
    \includegraphics[width=0.92\linewidth]{figs/Flowchart_architecture_Work.pdf}
    \caption{(a) Graphic representation of \elagenteS{}'s hierarchical architecture. At the top of the hierarchy is the \texttt{Computational Chemist Agent} (green node), responsible for planning calculation workflows based on a user's request. It is then assisted by four main subagents (nodes connected by solid and dashed arrows) responsible for accomplishing different parts of a calculation workflow: the \texttt{Geometry Generator Subagent}, which generates initial geometries; the \texttt{DFT Subagent}, which creates input files and runs \QE{} and \texttt{Phonopy}; the \texttt{File I/O Subagent}, which creates folders and moves files; and the \texttt{Output Analyzer Subagent}, which analyzes the output of a calculation, the star symbol indicates a connector in the flowchart. (b) Flowchart that outlines a typical calculation workflow executed by \elagenteS{}, the diamond symbol represents a connector in the flowchart.}
    \label{fig: flowchart_architecture}
\end{figure}

\subsection{Workflow Example}
In Figure \ref{fig: flowchart_architecture} (b), we illustrate a typical workflow executed by \elagenteS{}. At the start of the workflow, the \texttt{Computational Chemist Subagent} plans the overall procedure in response to the user’s query. The \texttt{Geometry Generator Subagent} then generates the initial geometries for all materials involved in the workflow and relaxes them using \uma{}. Next, the \texttt{DFT Agent} oversees the generation of \QE{} input files for each material and invokes the \texttt{\qe{} Running Agent} to perform the calculations. If the calculations complete successfully, the \texttt{Output Analyzer Subagent} analyzes the resulting outputs. If any calculation fails, the \texttt{DFT Subagent} troubleshoots the issue by modifying the input files and passing the revised files to the \texttt{\qe{} Running Subagent} for resubmission. This process is repeated for each step of the computational workflow until the final stage is reached. Once the final calculation is complete, the \texttt{Computational Chemist Subagent} prepares a final report summarizing the results

\section{Results}

We benchmarked \elagenteS{} across 11 research questions designed to evaluate its autonomous problem-solving capabilities in materials modelling using \texttt{\QE} and MLIPs. In the first seven exercises, we tested \elagenteS{}'s ability to answer foundational computational chemistry benchmark questions. These questions were designed to compute various properties across a diverse range of materials. For each benchmarking exercise, we formulated two versions of the same question with varying difficulty levels (\texttt{Level 1} and \texttt{Level 2}), each repeated five times (ten trials in total). We then evaluated each iteration using a rubric to assess the extent to which \elagenteS{} performed. These exercises focus on calculating fundamental material properties, such as band structures, bulk moduli, and relaxed geometries. The remaining four exercises are case studies that showcase \elagenteS{}'s ability to set up and execute complex computational materials workflows. These exercises are explicitly oriented toward materials discovery in areas including Li-ion battery materials, catalysis, and porous materials such as MOFs and COFs.

\subsection{Benchmarking Exercises}
Benchmarking exercises aim to assess convergence in energy cutoffs, bulk modulus, thermodynamic stability, surface energies, doping energies, and band structure. \texttt{Level 1} and \texttt{level 2} variants were included for most exercises. \texttt{Level 1} variants of the question typically have details that would greatly assist in answering each question; \texttt{level 2} variants do not. Table \ref{tab:seven_rubrics} summarizes the exercises, evaluation rubrics, and overall performance.
\setcounter{topnumber}{10}
\setcounter{dbltopnumber}{10}
\renewcommand{\topfraction}{1.0}
\renewcommand{\dbltopfraction}{1.0}
\renewcommand{\textfraction}{0.0}
\renewcommand{\floatpagefraction}{0.0}
\renewcommand{\dblfloatpagefraction}{0.0}
\setlength{\textfloatsep}{6pt plus 2pt minus 2pt}
\setlength{\floatsep}{4pt plus 2pt minus 2pt}
\setlength{\dbltextfloatsep}{6pt plus 2pt minus 2pt}
\setlength{\dblfloatsep}{4pt plus 2pt minus 2pt}

\newcommand{\RubLabel}[1]{%
  \par\noindent{\scriptsize\textbf{#1}}\par\vspace{0.1em}%
}

\newcommand{\RubricSubtable}[5]{%
\par\noindent
\begin{minipage}[htbp]{#1}
\centering
\begin{adjustbox}{max width=\linewidth, max totalheight=0.90\textheight, center}
\begin{minipage}{\linewidth}
\footnotesize
\setlength{\tabcolsep}{3pt}
\renewcommand{\arraystretch}{1.00}
\setlength{\aboverulesep}{2pt}
\setlength{\belowrulesep}{2pt}
\setlength{\cmidrulesep}{0pt}
\begin{tabularx}{\linewidth}{@{}p{2.9cm}p{0.5cm}X@{}}
\toprule
\multicolumn{3}{@{}>{\raggedright\arraybackslash}p{\dimexpr\linewidth\relax}@{}}%
{\textbf{Description:} #4}\\
\midrule
#5
\bottomrule
\end{tabularx}
\end{minipage}
\end{adjustbox}
\end{minipage}%
\par
}

\begin{table*}[b]
\caption{Evaluation rubrics for seven benchmarks calculating the following: (A) convergence testing, (B) stability, (C) bulk modulus, (D) surface energy, (E) doping energy, (F) structural relaxations, and (G) band structure.}
\label{tab:seven_rubrics}

\textbf{Exercise A}
\RubricSubtable{\textwidth}{a}{SCF convergence testing of $\alpha$-Fe}
{Perform convergence testing of $E_\text{cut}^{\text{wfc}}$ from 40 to 140 Ry (step 10 Ry) on $\alpha$-Fe to determine (i) the cutoff where the SCF energy error is below 10 meV/atom and (ii) the SCF energy error at $E_\text{cut}^{\text{wfc}}=100$ Ry. In Level 1 iterations, the $\alpha$-Fe structure is provided; in Level 2 iterations, the subagent must query $\alpha$-Fe from OQMD.}
{%
Planning & 30 & Only \texttt{calculation='scf'}. Set up calculations from \texttt{ecutwfc=40} to \texttt{140} Ry (step 10 Ry) (10). Keep inputs besides \texttt{ecutwfc} and \texttt{ecutrho} identical (10). \\
Geometry generation & 10 & For Level 1: use provided $\alpha$-Fe structure; for Level 2: query Fe from OQMD. (10) \\
Input file generation & 30 & Enforce \texttt{ecutrho} $\ge 8 \times$ \texttt{ecutwfc}. (10) Use Marzari--Vanderbilt smearing. (10) Set \texttt{starting\_magnetization} for Fe. (10) \\
Answer & 30 & (i) 90 Ry (10) (ii) 1.95 meV/atom (10) (iii) Final graph must be present. (10) \\
\midrule
\multicolumn{2}{@{}l}{\textbf{Difficulty level}} & \textbf{Average score} \\
\multicolumn{2}{@{}l}{Level 1} & 100.0 \\
\multicolumn{2}{@{}l}{Leve1 2} & 98.0 \\
}
\end{table*}
 
\begin{table*}[!b]\ContinuedFloat
\textbf{Exercise B}
\RubricSubtable{\textwidth}{c}{Calculate Energy Above Convex Hull of CaTi$_2$O$_4$}
{Calculate the energy above the convex hull of CaTi$_2$O$_4$. There is only one difficulty level with 10 iterations for this question.}
{%
Planning & 30 & Must deduce competing phases, CaTiO$_3$ and TiO (10), and their proportions, 5/7 and 2/7 respectively (10). Must plan out vc-relax -> scf for each structure (10). \\
Geometry generation & 10 & Query all three structures from OQMD. \\
Input file generation & 30 & ecutwfc $\geq$ 90, ecutrho $\geq$ 720 (10), DFT-D3 with BJ damping (10), conv\_thr $\leq$ 1E-6, etot\_conv\_thr $\leq$ 1E-5 (10). \\
Answer & 30 & 9 $\pm$ 3 meV/atom. \\
\midrule
\multicolumn{2}{@{}l}{\textbf{Difficulty level}} & \textbf{Average score} \\
\multicolumn{2}{@{}l}{Level 1} & 96.0 \\
\multicolumn{2}{@{}l}{Level 2} & N/A \\
}
\end{table*}
\begin{table*}[!b]\ContinuedFloat
\textbf{Exercise C}
\RubricSubtable{\textwidth}{e}{Calculate bulk modulus of Cu, MgO and Si}
{Calculate the bulk modulus of Cu, MgO, and Si using volumes from -7\% to +7\% with +1\% increments and plot an energy-volume curve. In Level 1 iterations, workflow details are provided, while in Level 2 iterations, they are not.}
{%
Planning & 20 & Must ensure use of 'relax', not 'vc-relax' (10). Must plot volume-energy curve and use this to calculate bulk modulus (10). \\
Geometry generation & 20 & Must query Si, MgO, and Cu from OQMD (10). Must relax with \uma{} and create +7\% to -7\% structures (10). \\
Input file generation & 30 & Marzari-Vanderbilt smearing for Cu, 'gaussian' smearing for Si and MgO (10). ecutwfc $\geq$ 90, ecutrho $\geq$ 720 (10). DFT-D3 with BJ damping (4). forc\_conv\_thr $\leq$ 5E-4 (2), etot\_conv\_thr $\leq$ 1E-5 (2), conv\_thr $\leq$ 1E-6 (2). \\
Answer & 30 & 162 $\pm$ 10 GPa for MgO (10), 161 $\pm$ 10 GPa for Cu (10), 91 $\pm$ 5 GPa for Si (10). \\
\midrule
\multicolumn{2}{@{}l}{\textbf{Difficulty level}} & \textbf{Average score} \\
\multicolumn{2}{@{}l}{Level 1} & 100.0 \\
\multicolumn{2}{@{}l}{Level 2} & 97.2 \\
}
\end{table*}

\begin{table*}[!t]\ContinuedFloat
\textbf{Exercise D}
\RubricSubtable{\textwidth}{c}{}
{Calculate the XRD pattern of Si, Cu, Pt, and Fe. Using the Miller index corresponding to the most intense peak of the XRD spectrum of each metal, calculate the surface energy using \uma{}. In Level 1 iterations, the surface energy formula is provided; in Level 2 iterations, the formula is omitted.}
{%
Planning & 20 & Plans to relax bulk structures with \uma{}, generate XRD spectra, pick the Miller index, generate the slabs with 4 layers of atoms, and calculate the energies. \\
Geometry generation & 20 & Query all structures from OQMD. \\
Input file generation & 30 & Variable cell relaxation of bulk structures -> surface construction based on correct Miller index -> relaxation with atomic positions alone. \\
Answer & 30 & Si:\SI{5.46}{\electronvolt\per\square\angstrom}(111), Fe:\SI{2.85}{\electronvolt\per\square\angstrom}(110), Cu:\SI{3.62}{\electronvolt\per\square\angstrom}(111), Pt:\SI{3.96}{\electronvolt\per\square\angstrom}(111). \\
\midrule
\multicolumn{2}{@{}l}{\textbf{Difficulty level}} & \textbf{Average score} \\
\multicolumn{2}{@{}l}{Level 1} & 98.0 \\
\multicolumn{2}{@{}l}{Level 2} & 96.0 \\
}
\end{table*}

\begin{table*}[!t]\ContinuedFloat
\textbf{Exercise E}
\RubricSubtable{\textwidth}{d}{Calculate the P and B doping energy of Si}
{Using \uma{} for pre-relaxations, PBE for DFT relaxations, and R2SCAN for SCF calculations, calculate the energy to dope B and P into Si. In \textbf{Level 1}, the equation for calculating the doping energy is provided.}
{%
Planning & 20 & Must determine formula for doping P and B, and the structures needed (10). Must determine vc-relax -> scf for supercells and relax -> scf for molecules (10). \\
Geometry generation & 40 & Query conventional Si (10). Construct supercell (10). Dope B and P in supercell (10). Construct PH$_3$, H$_2$, and B$_2$H$_6$ (10). \\
Input file generation & 20 & For molecules: assume\_isolated = 'martyna-tuckerman' (2), Gamma $k$-point (2), vacuum space of 10 \AA\ (2), occupations = 'fixed' (2). For supercells: ecutwfc $\geq$ 90, ecutrho $\geq$ 360 (2), etot\_conv\_thr $\leq$ 1E-5 (2). Use norm-conserving pseudopotentials throughout (6). \\
Answer & 20 & P-doping: 0.25$\pm$0.07 eV. B-doping: 0.95$\pm$0.07 eV. \\
\midrule
\multicolumn{2}{@{}l}{\textbf{Difficulty level}} & \textbf{Average score} \\
\multicolumn{2}{@{}l}{Level 1} & 98.8 \\
\multicolumn{2}{@{}l}{Level 2} & 98.0 \\
}
\end{table*}

\begin{table*}[!t]\ContinuedFloat
\textbf{Exercise F}
\RubricSubtable{\textwidth}{b}{Relaxations of 7 structures}
{Perform variable cell-relaxations of Si, Fe, LaMnO$_3$, LiFePO$_4$, MnO, anatase-TiO$_2$ and rutile-TiO$_2$. Report energies, magnetizations (total and absolute), cell parameters, and volumes. For Level 1, structures are provided, and the MnO, LaMnO$_3$, and LiFePO$_4$ antiferromagnetic character is mentioned in the prompt.}
{%
Planning & 20 & Overall workflow of vc-relax followed by scf (10). Must recognize that MnO, LaMnO$_3$, and LiFePO$_4$ are antiferromagnetic and set up input files as such (10). \\
Geometry generation & 10 & For Level 1, the structures must be extracted from the provided JSON files; for Level 2, all structures must be correctly queried from OQMD (10). \\
Input file generation & 41 & Set \texttt{starting\_magnetization} for Fe, MnO, LaMnO$_3$ and LiFePO$_4$ (4). Marzari--Vanderbilt smearing for Fe (1), 'gaussian' smearing for the rest (6). ecutwfc $\geq$ 90, ecutrho $\geq$ 720 (10). U values set for MnO, LaMnO$_3$, and LiFePO$_4$ (10). DFT-D3 with BJ damping (5). forc\_conv\_thr $\leq$ 5E-4, conv\_thr $\leq$ 1E-6. \\
Answer & 29 & Answers in Table S4--S6 in Supplementary Information. \\
\midrule
\multicolumn{2}{@{}l}{\textbf{Difficulty level}} & \textbf{Average score} \\
\multicolumn{2}{@{}l}{Level 1} & 98.6 \\
\multicolumn{2}{@{}l}{Level 2} & 95.8 \\
}
\end{table*}
\clearpage
\begin{table*}[!t]\ContinuedFloat
\textbf{Exercise G}
\RubricSubtable{\textwidth}{f}{Calculate and Plot Bandstructure of 5 materials}
{Calculate and plot the PBE bandstructures (no SOC) of Si, Fe, MnO, ZnCo$_2$O$_4$, and CsPbI$_3$. In Level 1, the structures are provided; in Level 2, they are not.}
{%
Planning & 14 & Must do vc-relax -> scf -> bands (10). MnO (2) and ZnCo$_2$O$_4$ (2) must be antiferromagnetic. \\
Geometry generation & 10 & Query structures from OQMD or retrieve them from provided JSONs (10). \\
Input file generation & 46 & starting\_magnetization for Fe, MnO, ZnCo$_2$O$_4$ (10). Marzari-Vanderbilt smearing for Fe (1), 'gaussian' for the rest (4). U values for Mn in MnO (5) and Co in ZnCo$_2$O$_4$ (5). ecutwfc $\geq$ 90, ecutrho $\geq$ 720 (10). DFT-D3 with BJ damping (5). conv\_thr $\leq$ 1E-6 (2), etot\_conv\_thr $\leq$ 1E-5 (2), forc\_conv\_thr $\leq$ 5E-4 (2). \\
Answer & 30 & Fe: 0 eV (2), metallic (2). Si: 0.57$\pm$0.05 eV (2), indirect (2). MnO: 1.68$\pm$0.05 eV (2), indirect (2). ZnCo$_2$O$_4$: 2.23$\pm$0.05 eV (2), indirect (2). CsPbI$_3$: 2.55$\pm$0.05 eV (2), direct (2). All plots should be saved (10). \\
\midrule
\multicolumn{2}{@{}l}{\textbf{Difficulty level}} & \textbf{Average score} \\
\multicolumn{2}{@{}l}{Level 1} & 99.6 \\
\multicolumn{2}{@{}l}{Level 2} & 98.4 \\
}
\end{table*}

Each benchmarking exercise required \elagenteS{} to independently select appropriate computational parameters, such as energy cutoffs, $k$-point meshes, density functionals (PBE/PBE+U/r2SCAN), and pseudopotentials, and to generate comprehensive reports detailing methodology, input parameters, and final results. All benchmarking exercises were performed in fully autonomous mode, with a single user prompt to facilitate the evaluation. We then compared the results obtained in each exercise with reference calculations performed independently by computational chemists.

The evaluation rubrics were primarily structured so that 70\% of the total grade was based on the procedure used by \elagenteS{} to fulfill the user’s request, while the remaining 30\% depended on the final answer computed by \elagenteS{}. A notable exception is the doping energy question in \texttt{Exercise E}, where, due to the complexity of the workflow, 80\% of the grade was assigned to the procedure.

The procedural grade was further divided into three components: (i) planning, in which marks were awarded based on the high-level strategy developed by \elagenteS{} to fulfill the user’s request; (ii) geometry generation, in which marks were awarded based on the procedure used to construct the initial geometries; and (iii) input file generation, in which marks were awarded based on the correctness of specific settings in the input files required to obtain the final result.

The average score across both \texttt{level 1} and \texttt{level 2} iterations is reported in Table 1. \elagenteS{} maintains an overall average score above 95\% across all questions and difficulty levels, achieving a 97.9\% average success rate, demonstrating its strong ability to perform solid-state calculations. A detailed analysis of each iteration is provided in the Supplementary Information.

\subsection{Case studies}

\subsubsection{Electrocatalytic activity}

One of the fundamental aspects governing catalytic activity in heterogeneous catalysts is the analysis of formation energies along equilibrium points on the potential energy surface. These points define the stable configurations of adsorbed molecules and are commonly regarded as reaction intermediates. The nature and stability of these intermediates can change when charge transfer occurs between the catalyst and the adsorbed species. In electrochemical systems, charge transfer requires overcoming an electrical potential known as the activation potential, which enables electron-proton exchange at the interface. In the case of the oxygen evolution reaction (OER), where oxygen-based species are transformed into water or hydrogen peroxide, a sequence of elementary steps involving coupled proton–electron transfers is required. The key quantities determining catalytic activity are the adsorption free energies of oxygen-containing intermediates ($^*$\ce{OH}, $^*$\ce{O}, and $^*$\ce{OOH}) on the catalyst surface.

Within this framework, surface thermodynamics are typically described by combining total energies obtained from DFT calculations with appropriate corrections accounting for zero-point energy, entropy, and electrochemical driving forces. The reaction free energies of individual OER steps are determined by referencing proton-electron transfers to a standard electrochemical potential, enabling direct comparison across different catalysts and surface terminations.

The computational hydrogen electrode (CHE) model provides a widely adopted and computationally efficient framework for this purpose~\cite{norskov_origin_2004}. In this approach, the chemical potential of a proton–electron pair is equal to half the chemical potential of molecular hydrogen, with an explicit dependence on the applied electrode potential. Within the CHE formalism, the free energy change of an elementary electrochemical step is expressed as

\begin{equation}
\Delta G = \Delta E_{\mathrm{DFT}} + \Delta \mathrm{ZPE} - T \Delta S - eU,
\end{equation}

where $\Delta E_{\mathrm{DFT}}$ is the DFT reaction energy, $\Delta \mathrm{ZPE}$ and $\Delta S$ are the zero-point energy and entropy changes, respectively, and $U$ is the applied electrode potential relative to the reversible hydrogen electrode.

\begin{figure*}[!h]
    \centering
    \includegraphics[width=0.98\textwidth]{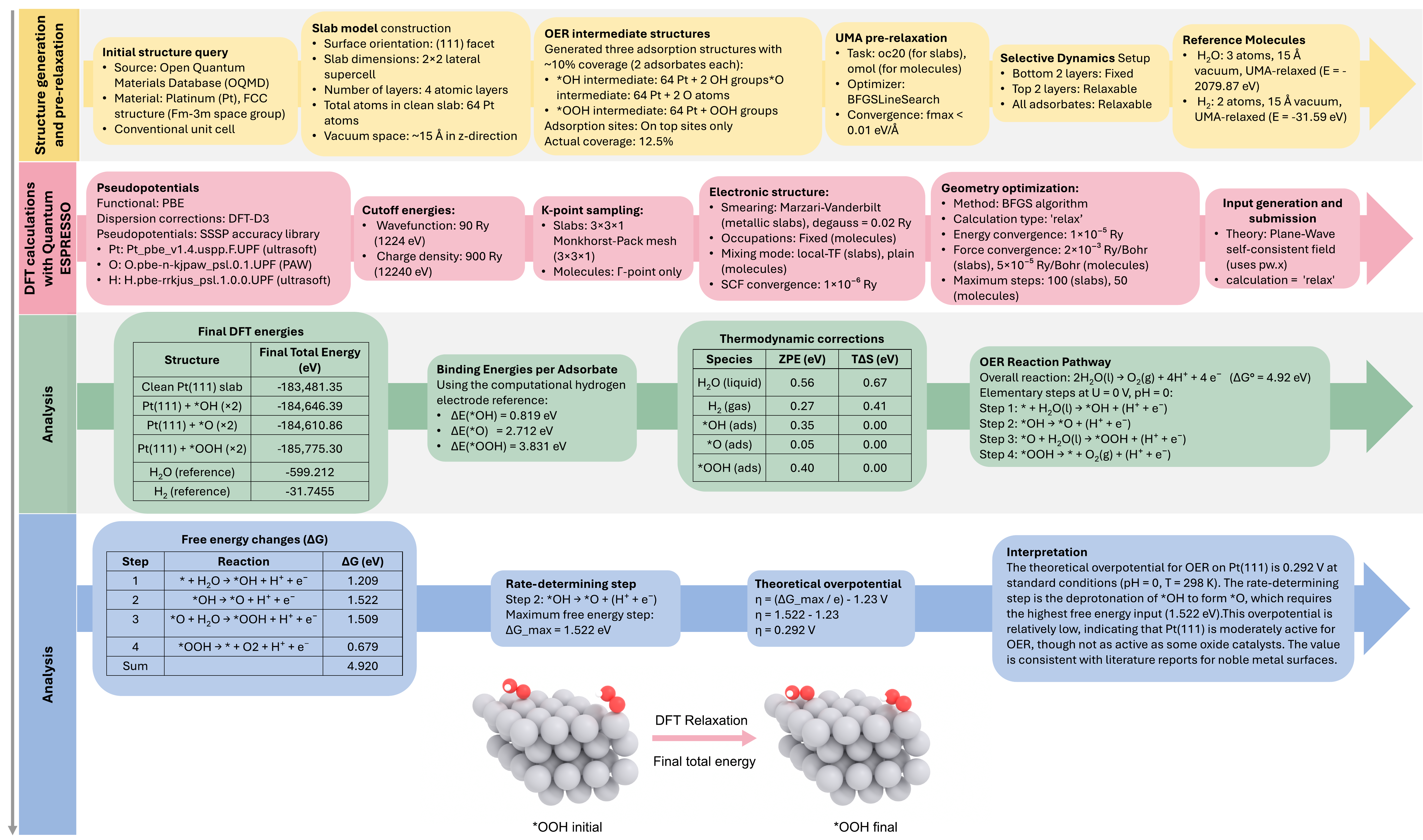}
    \caption{Flowchart summarizing the complete workflow used to calculate the theoretical OER overpotential on the Pt(111) surface.}
    \label{fig: Flowchart_OER}
\end{figure*}

\textbf{What this showcases}. This example demonstrates the agent’s ability to perform a full first-principles electrochemical catalysis workflow, from surface slab and adsorbate modelling to DFT-based reaction energetics within the computational hydrogen electrode framework. It demonstrates the automated calculation of adsorption free energies, the construction of OER free-energy diagrams, the identification of rate-determining steps, and the prediction of theoretical overpotentials using physically consistent thermodynamic corrections. Figure \ref{fig: Flowchart_OER} shows the workflow for this case study, together with the resulting outputs. Figures S4-S7 in the Supplementary Information showcase snippets of the output file detailing each step \elagenteS{} took to come to the final answer.

\prompt{You will execute a complete workflow to calculate the theoretical overpotential of OER on the surface of Pt with the computational hydrogen electrode model. You will first query the conventional unit cell of Pt from OQMD, and then turn that into a $2 \times 2$  supercell slab with 4 layers of atoms using the (111) surface. You will then create adsorption slab structures of OER intermediates on on-top sites with 10\% surface site coverage. Once that is done, you will relax them with \uma{}. Using the \uma{} relaxed structures, you will prepare selective dynamics and execute a complete DFT workflow to calculate the theoretical overpotential for OER at the Pt(111) surface. Remember to include any ZPE and entropy corrections using well-known values from the literature.}

The calculated OER free-energy diagram for the Pt(111) surface indicates that the deprotonation of $^*$\ce{OH} to form $^*$\ce{O} is the rate-determining step, with a maximum free-energy change of $\approx1.52$ eV at zero applied potential. This corresponds to a theoretical overpotential of $\approx0.29$ V, reflecting moderate catalytic activity under standard conditions. The relatively strong binding of oxygenated intermediates on Pt(111) stabilizes the initial reaction steps but energetically penalizes the formation of $^*$\ce{O}, consistent with known OER scaling relations. Overall, the agreement with prior theoretical and experimental studies supports the reliability of the computational setup and the suitability of the computational hydrogen electrode framework for describing the energetics of electrochemical reactions on metal surfaces \cite{iizuka2022tailoring}.

\subsubsection{Thermal properties}

\begin{figure}[h]
    \centering
    \includegraphics[width=0.85\textwidth]{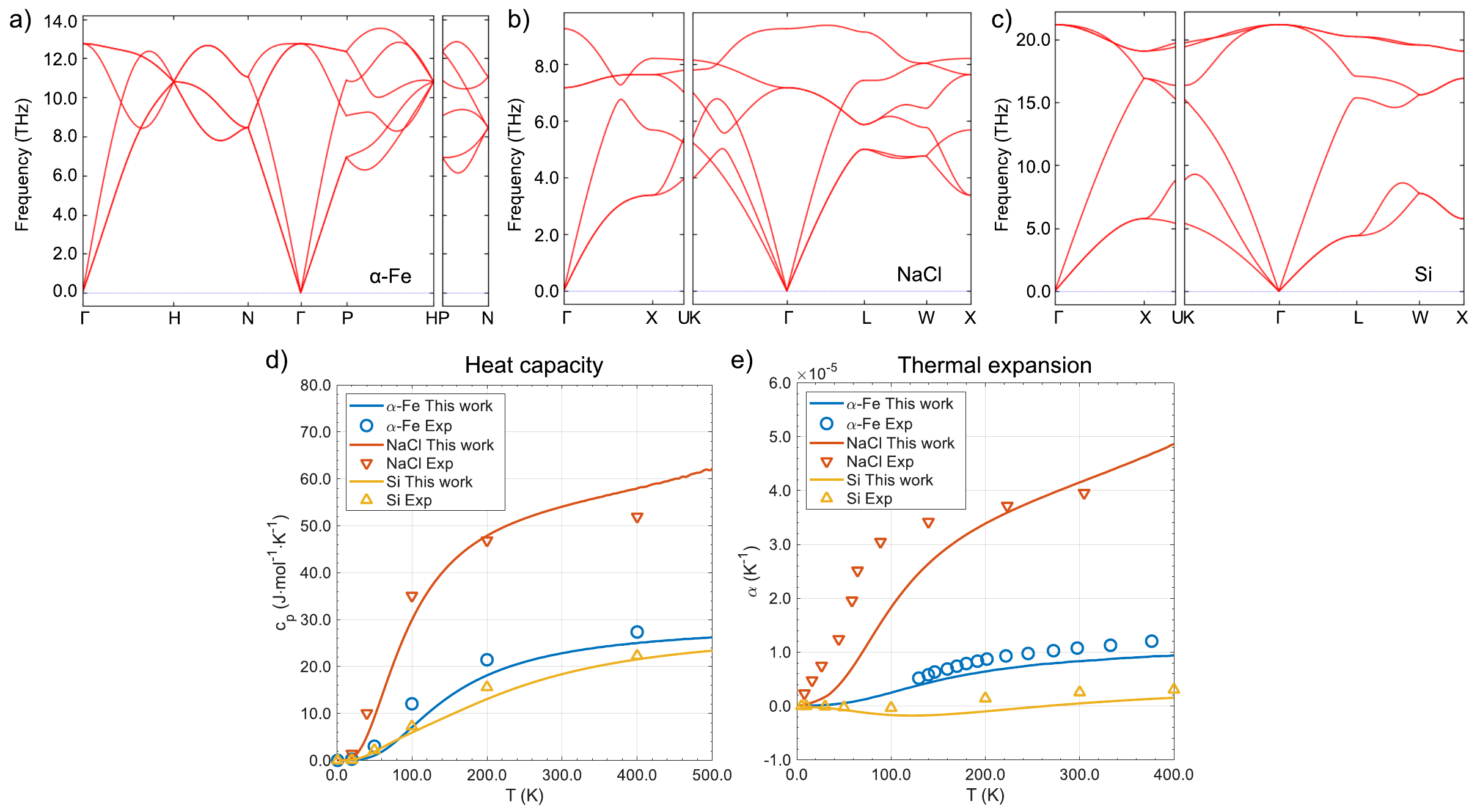}
    \caption{ Panels (a–c) show the phonon band structures of $\alpha$-Fe, NaCl, and Si at equilibrium volume, respectively. Panel (d) refers to the change in heat capacity of Fe, NaCl and Si with temperature at constant pressure and (e) refers to the thermal expansion coefficient of Fe, Si and NaCl. Plots a-c were generated by \elagenteS{}.
    Plots d and e were modified from plots generated by \elagenteS{} to include experimental thermal expansion coefficients for Si \cite{okada1984precise}, $\alpha$-Fe \cite{kozlovskii2019linear}, and NaCl \cite{pashaev2013epitaxial} as a reference. Figures S11-S25 in the Supplementary Information showcase all plots generated by \elagenteS{} when using \texttt{Phonopy} for QHA analysis.}
    \label{fig: Thermal_case}
\end{figure}

The finite-temperature thermodynamic properties of crystalline solids are governed by lattice vibrations and their response to volume changes. First-principles lattice-dynamical methods based on DFT enable the prediction of vibrational spectra and thermodynamic quantities without empirical parameters. Phonon calculations provide access to vibrational free energies, entropies, and heat capacities, while the volume dependence of phonon frequencies leads to thermal expansion and constant-pressure properties.

Among available approaches, the quasi-harmonic approximation (QHA) offers an efficient treatment of finite-temperature effects by allowing harmonic phonon frequencies to depend on volume. The Helmholtz free energy is written as

\begin{equation}
F(V,T) = E_{\mathrm{DFT}}(V) + F_{\mathrm{vib}}(V,T),
\end{equation}

where $E_{\mathrm{DFT}}(V)$ is the static DFT energy and $F_{\mathrm{vib}}(V,T)$ is the vibrational free energy evaluated from phonon frequencies at fixed volume. Minimization of $F(V,T)$ with respect to volume at each temperature yields the equilibrium volume and enables the calculation of thermal expansion coefficients, bulk moduli, and heat capacities at constant pressure. Minimizing this energy with respect to volume yields equilibrium volumes and derived thermodynamic quantities. By capturing the dominant anharmonic contribution from phonon–strain coupling while neglecting explicit phonon–phonon interactions, the QHA is accurate for dynamically stable materials at temperatures well below melting or phase transitions and has been successfully applied to a wide range of solids.

\textbf{What this showcases.} This example showcases the agent's ability to retrieve crystal structures, select appropriate first-principles methods and computational parameters, construct supercells, and orchestrate DFT calculations using \texttt{\QE} in combination with \texttt{Phonopy} for lattice-dynamical and quasi-harmonic analyses. It further demonstrates automated handling of metallic, covalent, and ionic materials, including the application of non-analytical corrections, and the generation of physically consistent phonon spectra and thermodynamic properties. Figure \ref{fig: Thermal_case} shows the outputs of this case study.

\prompt{Compute phonon dispersion relations for these crystalline solids: $\alpha$-Fe (body-centred cubic, ferromagnetic), Si (diamond cubic), and NaCl (rock-salt structure). Obtain the heat capacity, thermal conductivity, and Helmholtz free energy as a function of volume at standard conditions (298 K and 1 atm).}

The calculated phonon band structures reflect the distinct bonding character and lattice dynamics of the three materials considered. Silicon exhibits high-frequency optical modes and a large separation between acoustic and optical branches, consistent with its stiff covalent bonding network and relatively low atomic mass. The absence of imaginary frequencies across the sampled volume range confirms the dynamical stability of the diamond structure under both compressive and moderate tensile strains.

Across the three cases considered, the overall trend in thermal expansion is consistently reproduced by the experimental data; however, the agreement is less pronounced than for the heat capacity. This behaviour can be attributed to the use of the PBE pseudopotential in all three cases, which is known to systematically overestimate lattice constants compared to experimental values. As a result, the thermal expansion curve may exhibit a noticeable shift, as observed in all three systems studied. Furthermore, these discrepancies may also be related to the choice of computational parameters, such as the supercell size and the $k$-point sampling. In this regard, a $3 \times 3 \times 3$ supercell could provide a more accurate description; however, this option is computationally more demanding and therefore was not adopted in the present study. In the case of $\alpha$-Fe, its magnetic nature implies that volumetric expansion is strongly influenced by magnetic effects and possible magnon--phonon couplings, which are not explicitly captured within the quasi-harmonic approximation (QHA).

The calculated heat capacity at constant pressure for NaCl, Si, and $\alpha$-Fe exhibits behaviour consistent with established thermodynamic trends. At low temperatures, Cp follows the expected Debye-like temperature dependence governed by acoustic phonon modes, while at higher temperatures it approaches the classical Dulong–Petit limit. Differences among the three materials reflect variations in phonon spectra, bonding strength, and characteristic Debye temperatures.

At ambient conditions (298 K and 1 atm), the computed Cp values are in very good agreement with available experimental data for all three systems. This agreement demonstrates the robustness of the computational procedure and confirms the reliability of the combined DFT, \texttt{Phonopy}, and quasi-harmonic framework for predicting vibrational thermodynamic properties across metallic, covalent, and ionic materials.

The agreement between the calculated and experimental heat capacities further indicates that the underlying phonon dispersions accurately capture the relevant vibrational degrees of freedom, thereby validating their use in subsequent analyses of entropy, enthalpy, and free-energy contributions.hy Figures S11-S25 in the Supplementary Information contain all other plots generated by \elagenteS{} after performing the QHA analysis of NaCl, Si, and Fe using \texttt{Phonopy}.

\begin{figure}[b!]
    \centering
    \includegraphics[width=0.95\textwidth]{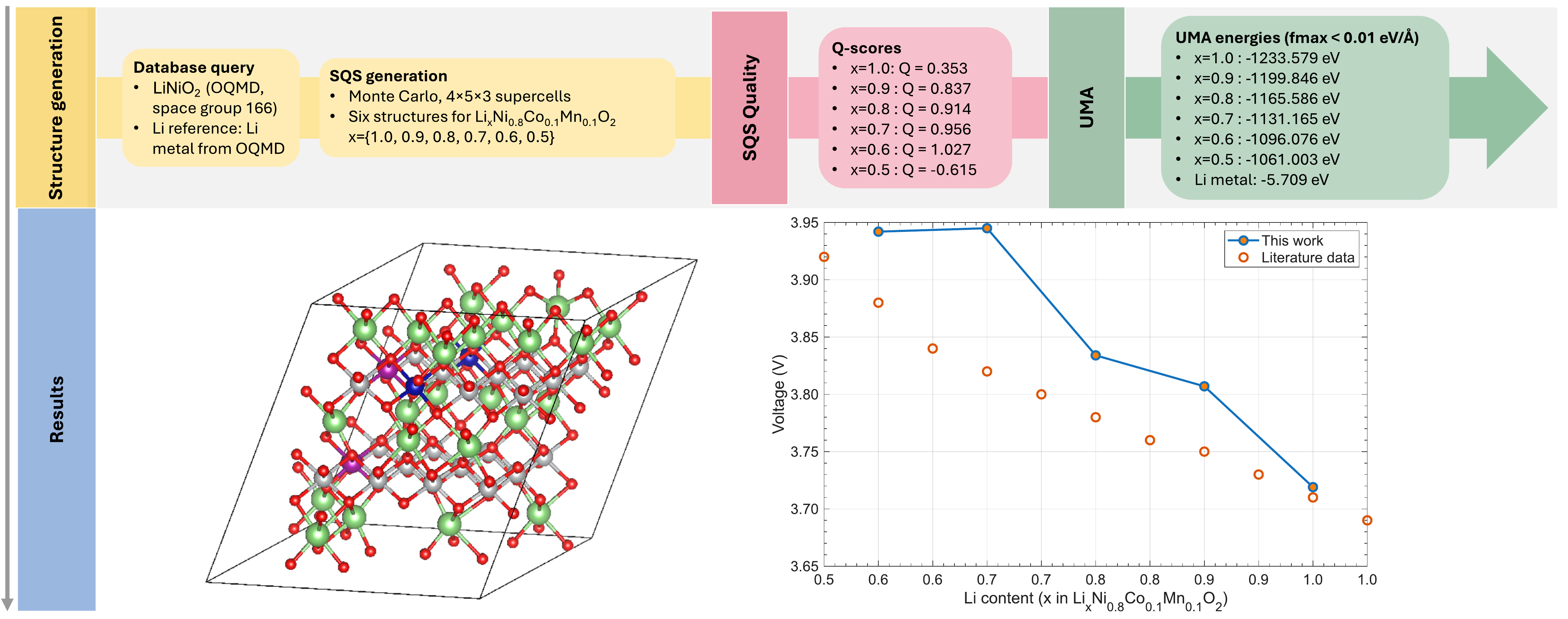}
    \caption{Flowchart summarizing the complete workflow used to compute the delithiation voltage profile of \ce{Li_{x}Ni_{0.8}Co_{0.1}Mn_{0.1}O_2} from x = 1.0 to 0.5 with intervals of 0.1. The delithiation voltages for experimental comparison were obtained from \cite{marker2019evolution}. The delithiation voltage plot was modified to include experimental data from a plot generated by \elagenteS{}. Figures S8-S10 contain details that \elagenteS{} took to generate the delithiated voltage curve of NMC-811.}
    \label{fig: Flowchart_lithium}
\end{figure}

\subsubsection{Electrochemical properties}

Delithiation voltages for layered \ce{Li_{x}Ni_{0.8}Co_{0.1}Mn_{0.1}O_2} where x, the stoichiometry of Li, is varied from 1.0 to 0.5,  were obtained using a computational workflow designed to reproduce the thermodynamic outputs of first-principles calculations without performing explicit DFT simulations for each configuration. Although plane-wave DFT codes are commonly employed for this purpose, their direct use becomes prohibitively expensive when sampling large supercells and multiple disordered lithium and vacancy configurations.

Instead, the present approach combines statistically representative structural models with pre-existing DFT knowledge and efficient surrogate relaxations. Configurational disorder was treated using special quasirandom structures (SQS) generated via Monte Carlo sampling, ensuring realistic lithium distributions across the compositional range of interest. Structural relaxations and energy evaluations were performed using a universal machine-learning–based interatomic model (\uma{}), which provides forces and energies at a fraction of the computational cost of conventional DFT while retaining physically meaningful trends.

Thermodynamic delithiation voltages were then computed from total energy differences using standard electrochemical expressions, with the lithium chemical potential referenced to bulk Li metal obtained from established first-principles databases. This strategy enables rapid evaluation of voltage profiles and compositional trends while preserving the essential physics captured by DFT, making it particularly suitable for high-throughput screening and exploratory studies of complex cathode materials.

\textbf{what this showcases.} This example showcases the agent’s ability to model delithiation in compositionally disordered cathode materials using special quasirandom structures (SQS) to represent lithium and vacancy disorder and a universal machine-learning–based interatomic model (\uma{}) for efficient structural relaxation and energy evaluation. It demonstrates how statistically representative configurations and surrogate potentials can be combined to compute thermodynamically consistent delithiation voltages without explicit DFT calculations, enabling rapid exploration of electrochemical trends in complex layered oxides. Figure \ref{fig: Flowchart_lithium} shows the workflow for this case study, together with the resulting outputs.  Figures S8-S10 in the Supplementary Information contain detailed snippets of the output file generated by \elagenteS{} that showcase the steps it took to generate the delithiated voltage curve of NMC-811.

\prompt{I want you to create a delithiation curve of \ce{Li_xNi_{0.8}Co_{0.1}Mn_{0.1}O_2} from 1 to 0.5 with decrements of 0.1 in x. You will first query the primitive cell of LiNiO$_2$ from OQMD with the space group number 166 (R-3m structure). Then you will use SQS to create 4x5x3 supercells of the primitive cell and generate all structures of \ce{Li_xNi_{0.8}Co_{0.1}Mn_{0.1}O_2} from 1 to 0.5 with a step size of 0.1 in x. Once this is complete, you will relax them with \uma{} and calculate their energies. Do not progress onto DFT. Create the delithiation curve using the \uma{} energies.
}

\subsubsection{MOF Construction and Bulk Modulus Calculation}

In this case study, we demonstrate \elagenteS{}'s ability to construct MOFs and COFs with \pormake{}, pre-relax the structures with \uma{}, and then subsequently calculate their bulk moduli using \QE. We use the pcu topology with the MOF and the dia topology with the COF in the following prompt:
\prompt{Using \pormake{}, I want you to construct a MOF and a COF for me. The MOF should be the simplest possible structure with pcu topology, nodes containing Zn and linkers containing terephthalamide. The COF should be the simplest possible structure with dia topology, without organic linkers. After constructing this MOF and COF, you will relax their structures with \uma{}. Then, you will calculate the bulk modulus of both structures with volumes from -6\% to +6\%, with +1\% increments by first pre-relaxing the positions of the atoms using \uma{}   for each structure and then using DFT with \QE{} to relax them(PBE functional).}

Upon receiving this request, the MOF Creator will first search \pormake{}'s database for nodes containing Zn and linkers containing terephthalimide that are compatible with the pcu topology. It will also look through \pormake{}'s database for the simplest possible organic node compatible with the dia topology. In the case of the MOF, the MOF Creator selected the Zn$_4$O node and a terephthalimide linker. In the case of the COF, the MOF Creator selected 2,2,5,5-tetramethylhexane as an appropriate node to construct it. Once the structures are constructed and relaxed with \uma{} and \elagenteS{}, the workflow performs a DFT calculation to determine the bulk moduli of both structures. 

\begin{figure*}[h]
    \centering
    \includegraphics[width=0.92\textwidth]{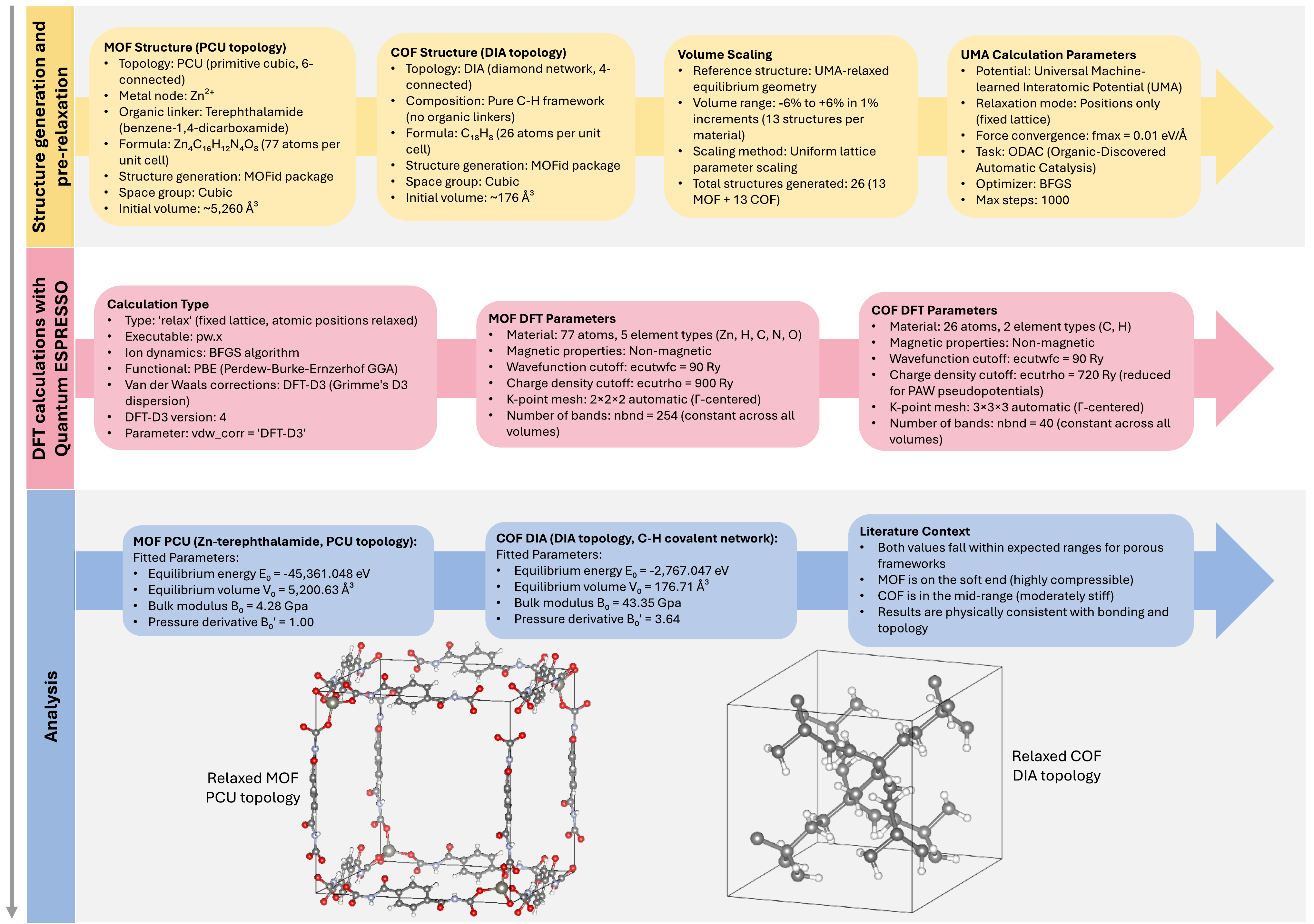}
    \caption{Computational workflow for bulk mechanical characterization of porous frameworks. The procedure involves initial structure construction, isotropic volume scaling, pre-relaxation using a machine-learned interatomic potential, and subsequent fixed-volume DFT relaxations. Structural renderings of the MOF and COF unit cells are shown for reference.}
    \label{fig:Flowchart_MOF}
\end{figure*}

\elagenteS{} successfully completed the 13 sets of calculations (pre-relaxation with UMA followed by variable-cell relaxation with \QE{}) required for each reticular framework. In its summary report, it produced a well-organized table containing the cell volume, total energy, and number of BFGS cycles needed to reach convergence extracted from the output files. From the extracted data, it plotted the energy-volume curves, and fitted the Birch-Murnaghan equation of state \cite{Birch_1947, Murnaghan_1944}, to the data. A summary of the calculation workflow and the results is shown in Figure \ref{fig:Flowchart_MOF}, and detailed snippets of the output file and the graphs generated by \elagenteS{} are found in Figures S24-S25 of the Supplementary Information.

\section{Conclusion}
In this work, we introduced \elagenteS{}, a hierarchical, LLM-driven multi-agent framework designed to autonomously plan, execute, and analyze first-principles solid-state simulations using \QE. \elagenteS{} enables researchers to specify high-level scientific objectives in natural language and obtain reproducible, end-to-end computational workflows with minimal manual intervention.

Through a comprehensive set of university-level benchmarking exercises, \elagenteS{} demonstrated robust and consistent performance across core tasks in solid-state computational chemistry, including convergence testing, geometry optimization, elastic property evaluation, surface energetics, electronic structure, and thermodynamic stability. Across two different levels of difficulty, the framework achieved a high average score of 97.9\% across all benchmarks, underscoring its ability to autonomously select appropriate physical models, numerical parameters, and workflow structures while adhering to established best practices.

Beyond controlled benchmarks, a series of realistic case studies highlighted the framework's flexibility and scientific reach. These included first-principles electrocatalysis workflows within the computational hydrogen electrode formalism, quasi-harmonic phonon calculations for finite-temperature thermodynamic properties, and data-efficient electrochemical voltage profiling using statistically representative disorder models and machine-learning interatomic potentials. Together, these examples illustrate how \elagenteS{} can seamlessly integrate DFT, lattice dynamics, surrogate models, and domain-specific physical formalisms within a unified autonomous pipeline.

More broadly, \elagenteS{} exemplifies a rapid shift toward agentic computational chemistry, in which expert knowledge is encoded not only in software packages but also in autonomous decision-making systems that reason about workflows, troubleshoot failures, and adapt strategies based on intermediate results. By lowering barriers to entry, enhancing reproducibility, and enabling scalable exploration of complex materials spaces, such systems have the potential to substantially accelerate materials discovery and design.

Looking forward, future developments will focus on expanding methodological coverage (e.g., excited-state methods, explicit solvation, and finite-field approaches), improving uncertainty quantification and validation strategies, and strengthening integration with experimental data and closed-loop discovery pipelines \cite{sdl_chemrev_2024, map_perspective_2023}. As large language models and foundation interatomic potentials continue to advance, frameworks such as \elagenteS{} point toward a future in which autonomous agents become routine collaborators in computational materials science rather than specialized tools used only by experts.

We look forward to releasing \elagenteS{} on our online platform under construction at \url{https://elagente.ca} .

\section{Data Availability}
All raw data pertaining to the benchmarking exercises and the case studies can be downloaded from \url{https://github.com/govlum/Benchmark_Results_El_Agente_Solido}



%% file: includes/include-acknowledgement.tex
We would like to dedicate this manuscript to Professor Geoffrey Ozin (University of Toronto). Aside from having contributed to many inspiring and thought-provoking discussions on the philosophical aspects of what AI will and will not do for science, he has been a great collaborator and mentor to many of us. 
J.V.H acknowledges the University of Toronto’s Acceleration Consortium, which receives funding from the CFREF-2022-00042 Canada First Research Excellence Fund. T.W.K. acknowledges the support of the Vector Distinguished Postdoctoral Fellowship. A. W. acknowledges support from NSERC through their Canada Graduate Scholarships - Doctoral (CGS-D) program as well as support from the Lawson Climate Institute.
A.A.-G. thanks Anders G. Fr{\o}seth, for his generous support. A.A.-G. also acknowledges the generous support of Natural Resources Canada and the Canada 150 Research Chairs program. This work was supported by the Defense Advanced Research Projects Agency (DARPA) under Agreement No. HR0011262E022. This work was also supported by the AI2050 program of Schmidt Sciences.
\acknowAC 
\acknowGEN{\emph{SciNet}}

%% file: includes/include-appendix.tex
\section{Supporting Information}
\label{app:related}
\begingroup
\setcounter{table}{0}
\renewcommand{\thetable}{S\arabic{table}}
\setcounter{figure}{0}
\renewcommand{\thefigure}{S\arabic{figure}}

\subsubsection{Summary of LLM Agentic Frameworks in Chemistry and Materials Science}
{
    \scriptsize  
    \setlength{\tabcolsep}{6pt} 
    \renewcommand{\arraystretch}{1.3} 
    \input{includes/lit_review_table}
}
\clearpage
\subsubsection{Evaluation of the Benchmark Question: Convergence Testing of ecutwfc for alpha-Fe}
Based on the rubrics, here are the detailed grades for all iterations we have done for this question:
\begin{table}[htbp]
\centering
\caption{Per-iteration grades for Level 1 vs Level 2 iterations.}
\label{tab:easy_hard_grades}

\begin{minipage}[t]{0.48\linewidth}
\centering
\textbf{Level 1 Iterations}\par\smallskip
\begin{tabular}{cc}
\toprule
Iteration & Grade \\
\midrule
1 & 100 \\
2 & 100 \\
3 & 100 \\
4 & 100 \\
5 & 100 \\
\bottomrule
\end{tabular}
\end{minipage}
\hfill
\begin{minipage}[t]{0.48\linewidth}
\centering
\textbf{Level 2 Iterations}\par\smallskip
\begin{tabular}{cc}
\toprule
Iteration & Grade \\
\midrule
1 & 100 \\
2 & 90 \\
3 & 100 \\
4 & 100 \\
5 & 100 \\
\bottomrule
\end{tabular}
\end{minipage}

\end{table}

For all Level 1 iterations, Solido achieved a perfect score. For the Level 2 iterations, Solido made an arithmetic mistake when calculating the error for 100Ry on one iteration, hence the deduction of 10 points. Solido achieved a perfect score in the other 4 iterations, thus giving it an average score of 98\% for the Level 2 iterations as well.

\subsubsection{Answers for the benchmark question: "Calculate the energy above the convex hull of CaTi$_2$O$_4$}

\renewcommand{\arraystretch}{1.2}
\begin{table}[htbp]
\centering
\caption{Per-iteration grades for the calculation of the energy above the convex hull of CaTi$_2$O$_4$}
\begin{tabular}{cc}
\textbf{Iteration} & \textbf{Grade} \\
\hline
1  & 100 \\
2  & 90  \\
3  & 100 \\
4  & 90  \\
5  & 100 \\
6  & 90  \\
7  & 100 \\
8  & 90  \\
9  & 100 \\
10 & 100 \\
\hline
\end{tabular}
\end{table}

For this question, there is only one difficulty level. 6 out of 10 iterations achieved a perfect score. For the ones that didn't, they all had one mistake in common: they did not mention that the proportions of the competing phases were 5/7 and 2/7 respectively in the summary output of the calculation. The fact that they had also arrived at the correct answer indicates that they had used correct proportions. Regardless marks were still deducted for not including this crucial detail in the output summary.
\clearpage
\subsubsection{Answers for the benchmark question: "Relax 7 structures"}
\FloatBarrier
\renewcommand{\arraystretch}{1.2}
\begin{table}[htbp]
\centering
\begin{threeparttable}
\caption{Reference energies (per atom / per formula unit as stated).}
\label{tab:energies}
\begin{tabular}{l l}
\toprule
\textbf{System} & \textbf{Energy Per Atom (eV/atom)}\\
\midrule
Fe                 & -4480.29 $\pm$ 0.10 \\
Si                 & -155.67  $\pm$ 0.10 \\
TiO\textsubscript{2} (anatase) & -920.74  $\pm$ 0.10\\
TiO\textsubscript{2} (rutile)  & -920.73  $\pm$ 0.10\\
MnO                & -1720.54 $\pm$ 0.10\\
LaMnO\textsubscript{3}         & -2163.07 $\pm$ 0.10\\
LiFePO\textsubscript{4}        & -1020.26 $\pm$ 0.10\\
\bottomrule
\end{tabular}
\begin{tablenotes}
\footnotesize
\item (1) Criterion / task reference as provided.
\end{tablenotes}
\end{threeparttable}
\end{table}

\begin{table}[htbp]
\centering
\begin{threeparttable}
\caption{Magnetization reference values.}
\label{tab:magnetization}

\begin{tabularx}{\linewidth}{@{}l p{0.24\linewidth} X@{}}
\toprule
\textbf{System} & \textbf{Total Magnetization} & \textbf{Absolute Magnetization} \\
\midrule

LiFePO\textsubscript{4} &
$0$ &
$15.71 \pm 0.10~\mu_B$/cell; $\sim 3.45 \pm 0.03~\mu_B$/Fe (4 Fe/cell) \\

MnO &
$0$ &
$9.49 \pm 0.05~\mu_B$/cell; $\sim 4.25 \pm 0.03~\mu_B$/Mn \\

LaMnO\textsubscript{3} &
$0$ &
$16.44 \pm 0.10~\mu_B$/cell; $\sim 3.29 \pm 0.03~\mu_B$/Mn (4 Mn/cell) \\

Fe (bcc) &
$4.24 \pm 0.05~\mu_B$/cell; $2.12 \pm 0.03~\mu_B$/Fe &
$4.45 \pm 0.05~\mu_B$/cell \\

\bottomrule
\end{tabularx}

\begin{tablenotes}
\footnotesize
\item Criterion / task reference as provided.
\item $\mu_B$ denotes Bohr magnetrons.
\end{tablenotes}

\end{threeparttable}
\end{table}

\begin{table}[htbp]
\centering
\begin{threeparttable}
\caption{Volumes and lattice parameters (with $\pm 1\%$ tolerance as stated).}
\label{tab:cellparams}
\begin{tabular}{l l l}
\toprule
\textbf{System} & \textbf{Volume (\AA$^3$)} & \textbf{Cell parameters (\AA)} \\
\midrule
Fe (bcc) &
$22.25 \pm 1\%$ &
$2.81 \times 2.81 \times 2.81 \; (\pm 1\%)$ \\

Si (diamond) &
$39.90 \pm 1\%$ &
$3.84 \times 3.84 \times 3.84 \; (\pm 1\%)$ \\

TiO\textsubscript{2} (anatase) &
$68.77 \pm 1\%$ &
$5.49 \times 5.49 \times 5.49 \; (\pm 1\%)$ \\

TiO\textsubscript{2} (rutile) &
$62.80 \pm 1\%$ &
$4.61 \times 4.61 \times 2.95 \; (\pm 1\%)$ \\

LaMnO\textsubscript{3} &
$253.9 \pm 1\%$ &
$5.53 \times 5.99 \times 7.66 \; (\pm 1\%)$ \\

LiFePO\textsubscript{4} &
$295.5 \pm 1\%$ &
$6.04 \times 4.72 \times 10.37 \; (\pm 1\%)$ \\

MnO\textsuperscript{a} &
$44.84 \pm 1\%$ &
$3.18 \times 3.18 \times 6.28 \; (\pm 1\%)$ \\
\bottomrule
\end{tabular}
\begin{tablenotes}
\footnotesize
\item Criterion / task reference as provided.
\item a. Stated for a $2\times1\times1$ supercell; normalize if using a different cell.
\end{tablenotes}
\end{threeparttable}
\end{table}

\subsubsection{Evaluation of the benchmark question: 'Relax 7 structures'}

\begin{table}[htbp]
\centering
\caption{Per-iteration grades for Level 1 vs Level 2 iterations.}
\label{tab:easy_hard_grades}

\begin{minipage}[t]{0.48\linewidth}
\centering
\textbf{Level 1 Iterations}\par\smallskip
\begin{tabular}{cc}
\toprule
Iteration & Grade \\
\midrule
1 & 99 \\
2 & 97 \\
3 & 98 \\
4 & 99 \\
5 & 99 \\
\bottomrule
\end{tabular}
\end{minipage}
\hfill
\begin{minipage}[t]{0.48\linewidth}
\centering
\textbf{Level 2 Iterations}\par\smallskip
\begin{tabular}{cc}
\toprule
Iteration & Grade \\
\midrule
1 & 96 \\
2 & 96 \\
3 & 95 \\
4 & 95 \\
5 & 97 \\
\bottomrule
\end{tabular}
\end{minipage}

\end{table}

We will be scrutinizing each run individually since the failure modes of each run are different. Let's start with the Level 1 Iterations:

Run 1: El Agente Solido set an ecutrho of 360 Ry for an ecutwfc of 90 Ry for one calculation (Si), when an ecutrho of at least 720 Ry should have been set. Since this mistake was only made for one structure, and the answers were well within the error range for the rest, this run is given 99 points.

Run 2: El Agente Solido set an ecutrho of 360 Ry for an ecutwfc of 90 Ry for one calculation (Si), when an ecutrho of at least 720 Ry should have been set. One point was deducted for this mistake. El Agente Solido also neglected to report cell parameters for both anatase-TiO$_2$ and MnO even though both were within an acceptable range, 2 points were deducted for this omission. Hence, this run has been given 97 points.

Run 3: When performing the 'scf' calculation for LiFePO$_4$, El Agente Solido mistakenly assigned a magnetic moment of 0 to the Fe atoms, and Fe's actual magnetic moment to another element. This caused the absolute magnetization to converge to 0. This error resulted in the final energy of LiFePO$_4$ to be off as well. There were no other mistakes in this run and the structures are generally accurate(since this mistake was only made in the 'scf' run). Every other structure was correct. Hence, this run was awarded 98 points.

Run 4: The calculated volume of LaMnO$_3$ is slightly outside the acceptable error range. Every other property has been accurately calculated for a final score of 99 points. 

Run 5: El Agente Solido made an arithmetic mistake when calculating the volume of LiFePO$_4$. This was the only mistake made and every other property has been accurately calculated, for a final score of 99 points.

Now we will be scrutinizing the 'hard' runs where the original structures were not provided:

Run 1: The calculated energies of anatase-TiO$_2$ and rutile TiO$_2$ were outside the acceptable error range, as is the volume and the cell parameters of anatase-TiO$_2$. This deviation from the correct answer is likely because El Agente Solido had erroneously assigned U Values for TiO$_2$ even though this is not typically done (in either Materials Project or OQMD). The rest of the structures were correct so we gave it a score of 96.

Run 2: When setting up the input file for LiFePO$_4$, El Agente Solido assigned magnetizations to the wrong atoms, thus causing the value of the magnetic moments of Fe to converge to 0. This caused the structure, the magnetic moments and the energy of LiFePO$_4$ to be incorrect. Every other structure is correct, thus giving this run a score of 96.

Run 3: El Agente Solido used an ecutrho value that is too low for Si (360 Ry) instead of it being 8 times ecutwfc. El Agente Solido also neglected to include van-der-Waals dispersion corrections for Fe. As a result the energies of both Si and Fe are off, as is the absolute magnetization of Fe. This is why we give this run a score of 95.

Run 4: Just like in Run 1, El Agente Solido assigned U values to both anatase TiO$_2$ and rutile TiO$_2$ causing the calculated energy to be off. The absolute magnetization of LaMnO$_3$ was slightly out of the error range for the magnetization of LaMnO$_3$. Hence, this run was given a score of 95.

Run 5: Total and absolute magnetizations of Fe and LiFePO$_4$ are incorrect. These were the only errors made by the agent, therefore this run was given a score of 97.

\subsubsection{Evaluation of the benchmark question "Calculate the PBE Bandstructure of Si, Fe, MnO, ZnCo$_2$O$_4$ and CsPbI$_3$}

\begin{table}[htbp]
\centering
\caption{Per-iteration grades for Level 1 vs Level 2 iterations.}
\label{tab:easy_hard_grades}

\begin{minipage}[t]{0.48\linewidth}
\centering
\textbf{Level 1 Iterations}\par\smallskip
\begin{tabular}{cc}
\toprule
Iteration & Grade \\
\midrule
1 & 98 \\
2 & 100 \\
3 & 100 \\
4 & 100 \\
5 & 100 \\
\bottomrule
\end{tabular}
\end{minipage}
\hfill
\begin{minipage}[t]{0.48\linewidth}
\centering
\textbf{Level 2 Iterations}\par\smallskip
\begin{tabular}{cc}
\toprule
Iteration & Grade \\
\midrule
1 & 96 \\
2 & 100 \\
3 & 100 \\
4 & 100 \\
5 & 96 \\
\bottomrule
\end{tabular}
\end{minipage}

\end{table}

 We shall first discuss the mistakes made in the Level 1 iterations without perfect scores:

 Run 1) This run did not include van-der-Waals dispersion corrections for the Fe calculation. This led to the deduction of 2 points.The rest of the runs have perfect scores

 We shall next discuss the mistakes made in the Level 2 iterations without perfect scores:

 Run 1) Points were deducted because ecutrho of Si was set to 360, 4 times that of ecutwfc, instead of eight times. Points were also deducted because CsPbI$_3$ was not recognized as a direct bandgap semiconductor.

 Run 5) Points were deducted because ZnCo$_2$O$_4$ was not recognized as being antiferromagnetic below the Neel temperature. Does not recognize that CsPbI3 has a direct band gap (reported as indirect)

\subsubsection{Evaluation of the benchmark question ``Bulk Modulus of Materials''}

\begin{table}[htbp]
\centering
\caption{Per-iteration grades for Level 1 vs Level 2 iterations.}
\label{tab:easy_hard_grades}

\begin{minipage}[t]{0.48\linewidth}
\centering
\textbf{Level 1 Iterations}\par\smallskip
\begin{tabular}{cc}
\toprule
Iteration & Grade \\
\midrule
1 & 100 \\
2 & 100 \\
3 & 100 \\
4 & 100 \\
5 & 100 \\
\bottomrule
\end{tabular}
\end{minipage}
\hfill
\begin{minipage}[t]{0.48\linewidth}
\centering
\textbf{Level 2 Iterations}\par\smallskip
\begin{tabular}{cc}
\toprule
Iteration & Grade \\
\midrule
1 & 96 \\
2 & 96 \\
3 & 96 \\
4 & 100 \\
5 & 96 \\
\bottomrule
\end{tabular}
\end{minipage}
\end{table}

All non-perfect runs have the same issue in common: they all suggest an ecutrho that is too low, less than 8 times that of the ecutwfc, for one of the materials. In these cases, an ecutrho of 450Ry was suggested for an ecutwfc of 90 Ry.

\subsubsection{Evaluation of the benchmark question: "Energy of doping P and B into Si"}

\begin{table}[htbp]
\centering
\caption{Per-iteration grades for Level 1 vs Level 2 iterations.}
\label{tab:easy_hard_grades}

\begin{minipage}[t]{0.48\linewidth}
\centering
\textbf{Level 1 Iterations}\par\smallskip
\begin{tabular}{cc}
\toprule
Iteration & Grade \\
\midrule
1 & 100 \\
2 & 96 \\
3 & 100 \\
4 & 98 \\
5 & 100 \\
\bottomrule
\end{tabular}
\end{minipage}
\hfill
\begin{minipage}[t]{0.48\linewidth}
\centering
\textbf{Level 2 Iterations}\par\smallskip
\begin{tabular}{cc}
\toprule
Iteration & Grade \\
\midrule
1 & 96 \\
2 & 98 \\
3 & 100 \\
4 & 96 \\
5 & 100 \\
\bottomrule
\end{tabular}
\end{minipage}
\end{table}

Now let's scrutinize the runs that did not get perfect scores, starting with the Level 1 iterations:

Run 2: Gamma k-point was not used for the calculation on molecules, neither were 'fixed' occupations. Because of these mistakes, 4 points were deducted for an overall score of 96.

Run 4: 'fixed' occupations were not used for the calculations on molecules. Because of this mistake, 2 points were deducted for an overall score of 96.

Now let's scruitinize the Level 2 iterations:

Run 1: B- doping energy was slightly outside acceptable range. Marks were still awarded for getting everything else right for a total score of 96.

Run 2: 2 points were taken off for not setting 'fixed' occupations for molecules, for an overall score of 98 points.

Run 4: P-doping energy was slightly outside acceptable range. Part marks awarded for getting everything else right for a total score of 96.

\subsubsection{Evaluation of the benchmark question ``Surface energies of elemental systems''}

\begin{table}[htbp]
\centering
\caption{Per-iteration grades for Level 1 vs Level 2 iterations.}
\label{tab:easy_hard_grades}

\begin{minipage}[t]{0.48\linewidth}
\centering
\textbf{Level 1 Iterations}\par\smallskip
\begin{tabular}{cc}
\toprule
Iteration & Grade \\
\midrule
1 & 100 \\
2 & 90 \\
3 & 100 \\
4 & 100 \\
5 & 100 \\
\bottomrule
\end{tabular}
\end{minipage}
\hfill
\begin{minipage}[t]{0.48\linewidth}
\centering
\textbf{Level 2 Iterations}\par\smallskip
\begin{tabular}{cc}
\toprule
Iteration & Grade \\
\midrule
1 & 100 \\
2 & 90 \\
3 & 100 \\
4 & 100 \\
5 & 90 \\
\bottomrule
\end{tabular}
\end{minipage}
\end{table}

Now let's scrutinize the runs that did not get perfect scores, starting with the Level 1 iterations:

Run 2: The agent successfully generates the correct crystal structures and performs variable-cell geometry relaxations. It also correctly constructs surface slabs based on the specified Miller indices derived from the XRD spectra. Finally, the agent relaxes the atomic positions and computes the surface energies using the correct formula. However, some slab relaxation input settings are not configured correctly.

Now let's scrutinize the Level 2 iterations:

Run 2: The agent successfully generates the correct crystal structures and performs variable-cell geometry relaxations. It also correctly constructs surface slabs based on the specified Miller indices derived from the XRD spectra. Finally, the agent relaxes the atomic positions and computes the surface energies using the correct formula. However, some slab relaxation input settings are not configured correctly.

Run 5: The agent successfully generates the correct crystal structures and performs variable-cell geometry relaxations. It also correctly constructs surface slabs based on the specified Miller indices derived from the XRD spectra. Finally, the agent relaxes the atomic positions and computes the surface energies using the correct formula. However, some slab relaxation input settings are not configured correctly.

\subsubsection{Electronic structure}

\begin{figure*}[htbp]
    \centering
    \includegraphics[width=0.98\textwidth]{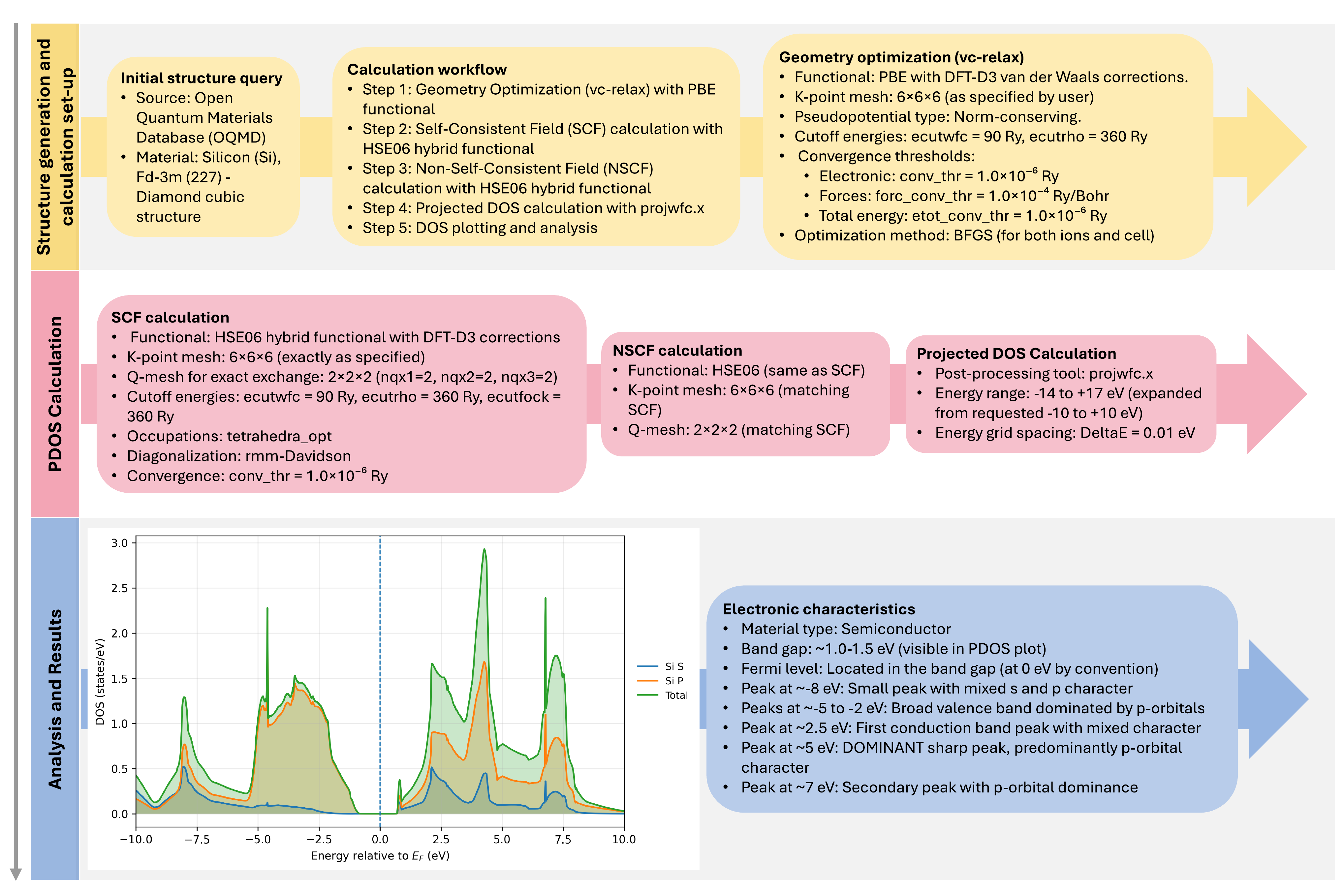}
    \caption{In this, we do an additional case study where we calculate the projected density-of-states of Si using the HSE functional}
    \label{fig:my_fig}
\end{figure*}

\clearpage
\subsection{Figures for OER case study}
\begin{figure*}[htbp]
    \centering
    \includegraphics[scale=0.75]{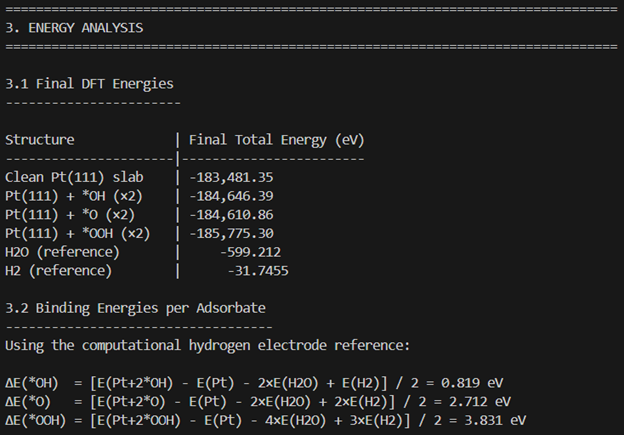}
    \caption{\elagenteS{} extracted the energy of all slabs and molecules, and performed some preliminary calculations of the energies of each step}
\end{figure*}

\begin{figure*}[htbp]
    \centering
    \includegraphics[scale=0.75]{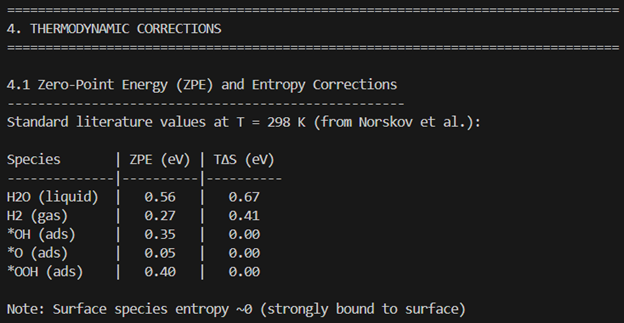}
    \caption{\elagenteS{} determines entropy and zero point energy (ZPE) corrections for each intermediate and molecule from the literature}
\end{figure*}

\begin{figure*}[htbp]
    \centering
    \includegraphics[scale=0.75]{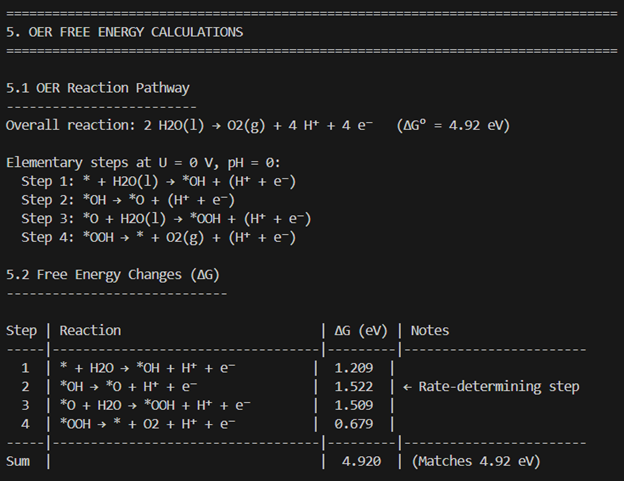}
    \caption{\elagenteS{} writes out all the steps of OER and the calculated energy of each step after applying ZPE and entropy corrections}
\end{figure*}

\begin{figure*}[htbp]
    \centering
    \includegraphics[scale=0.75]{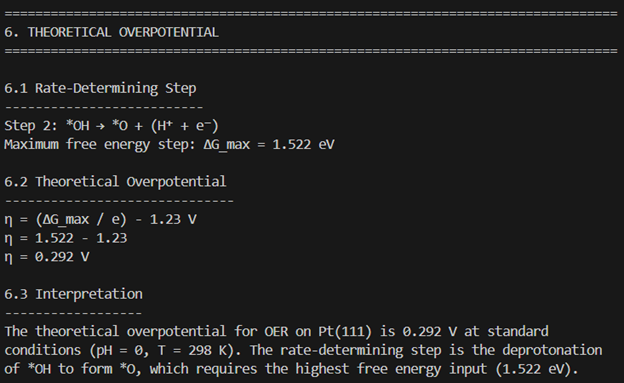}
    \caption{\elagenteS{} calculates the theoretical overpotential of OER on the surface of Pt.}
\end{figure*}
\FloatBarrier

\subsection{Figures for Li Ion Battery Delithiation Study}

\begin{figure*}[htbp]
    \centering
    \includegraphics[scale=0.75]{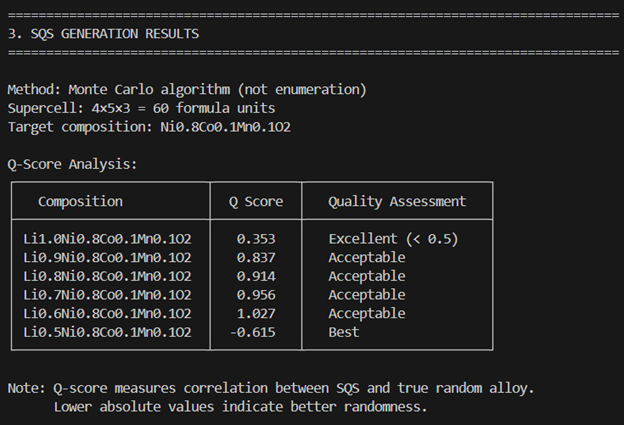}
    \caption{\elagenteS{} generates SQS structures of Li$_{x}$Ni$_{0.8}$Co$_{0.1}$Mn$_{0.1}$O$_2$ from x = 1.0 to x = 0.5}
\end{figure*}

\begin{figure*}[htbp]
    \centering
    \includegraphics[scale=0.75]{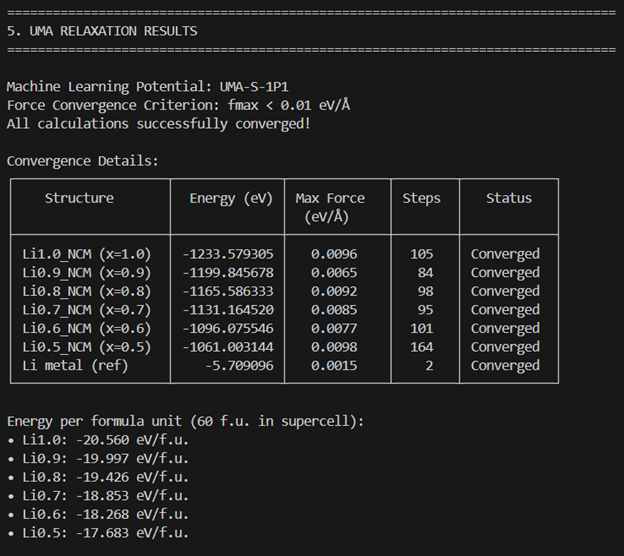}
    \caption{\elagenteS{} relaxes and calculates the energy of each SQS structure using UMA.}
\end{figure*}

\begin{figure*}[htbp]
    \centering
    \includegraphics[scale=0.75]{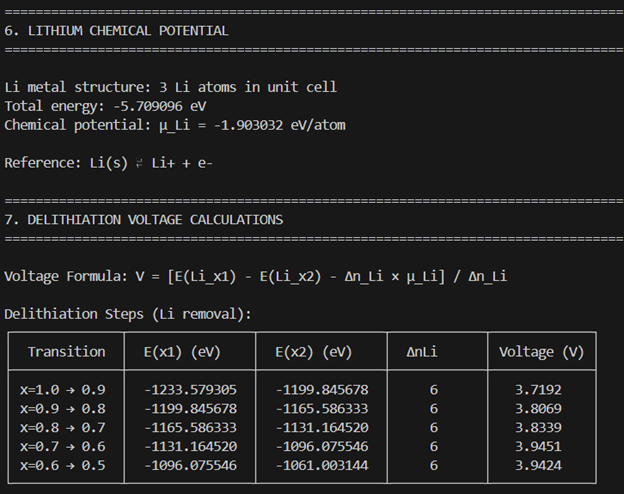}
    \caption{\elagenteS{} calculates the energy of Li metal using UMA and calculates the delithiation voltage at different stages of delithiation.}
\end{figure*}
\FloatBarrier

\subsection{Figures for the calculation of the thermal properties of Si, NaCl and Fe}

\begin{figure*}[htbp]
    \centering
    \includegraphics[scale=0.75]{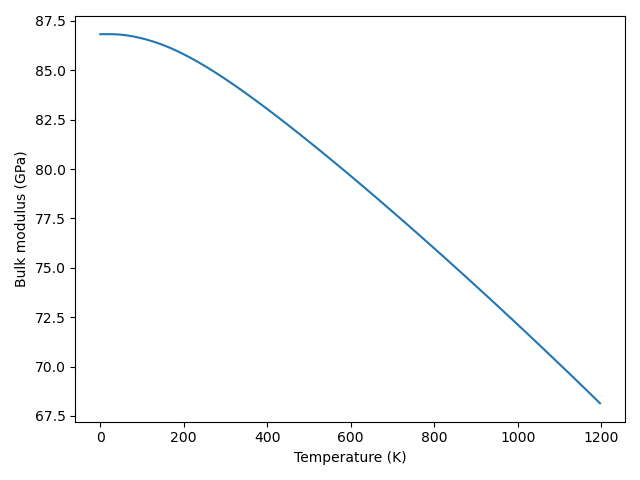}
    \caption{Change in bulk modulus of Si with temperature as calculated by \elagenteS{} with Phonopy. \elagenteS{} generated this plot.}
\end{figure*}

\begin{figure*}[htbp]
    \centering
    \includegraphics[scale=0.75]{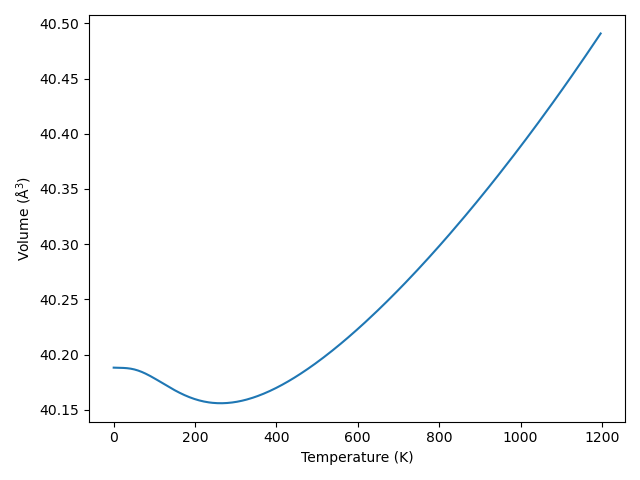}
    \caption{Change in volume of Si with temperature as calculated by \elagenteS{} with Phonopy. \elagenteS{} generated this plot.}
\end{figure*}

\begin{figure*}[htbp]
    \centering
    \includegraphics[scale=0.75]{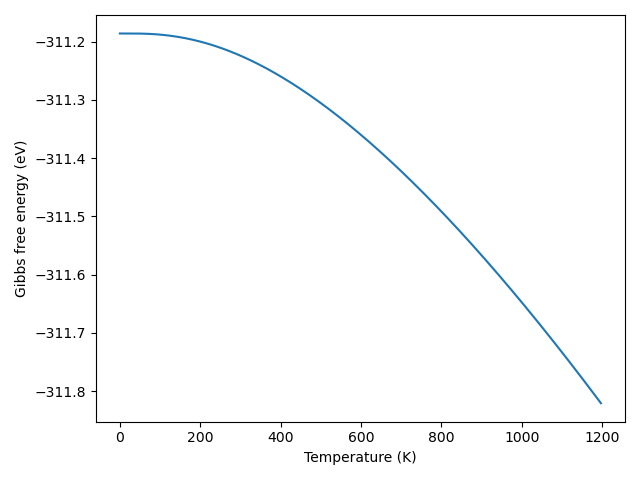}
    \caption{Change in Gibbs free energy of Si with temperature as calculated by \elagenteS{} with Phonopy. \elagenteS{} generated this plot.}
\end{figure*}

\begin{figure*}[htbp]
    \centering
    \includegraphics[scale=0.75]{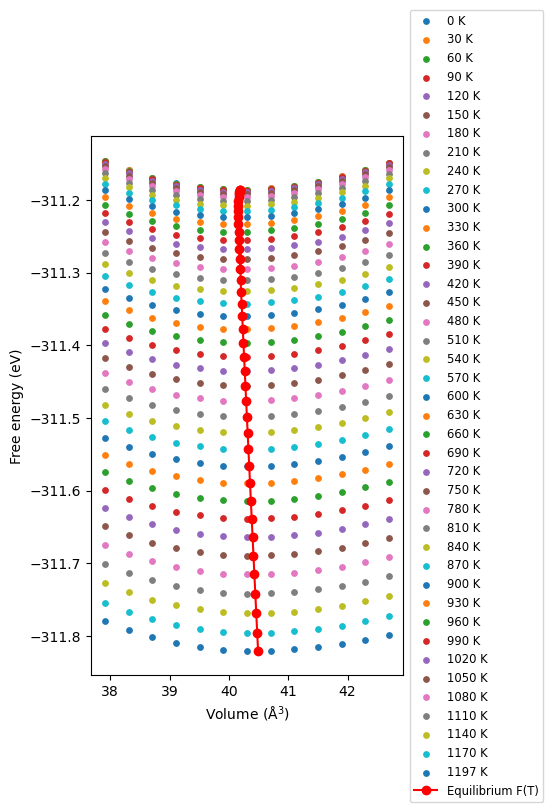}
    \caption{Change in Helmholtz free energy of Si with temperature and volume as calculated by \elagenteS{} with Phonopy. \elagenteS{} generated this plot.}
\end{figure*}

\begin{figure*}[htbp]
    \centering
    \includegraphics[scale=0.75]{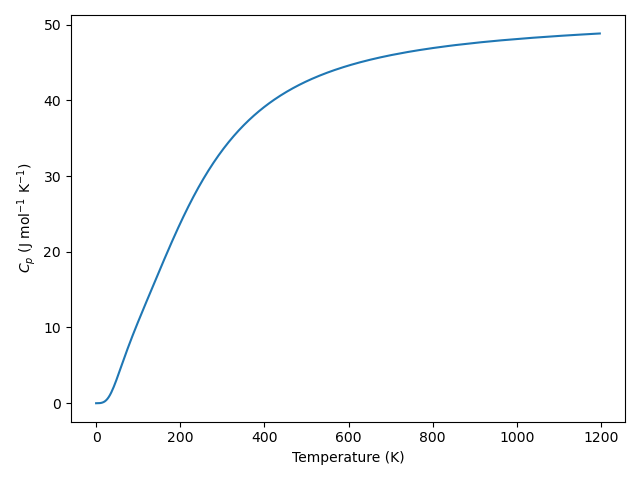}
    \caption{Change in heat capacity (C$_p$) of Si with temperature and volume as calculated by \elagenteS{} with Phonopy. \elagenteS{} generated this plot.}
\end{figure*}

\begin{figure*}[htbp]
    \centering
    \includegraphics[scale=0.75]{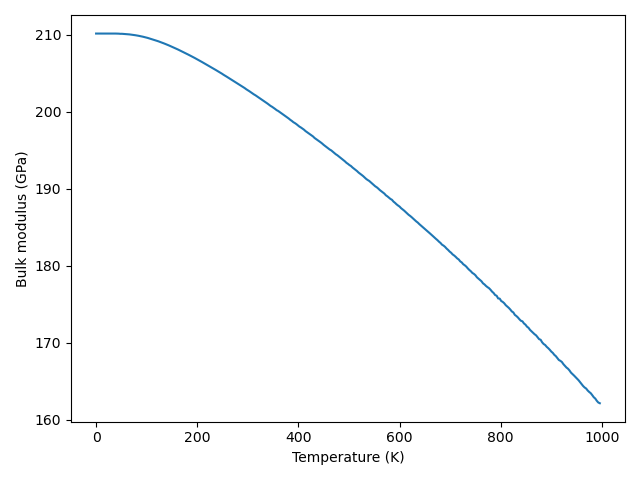}
    \caption{Change in bulk modulus of Fe with temperature as calculated by \elagenteS{} with Phonopy. \elagenteS{} generated this plot.}
\end{figure*}

\begin{figure*}[htbp]
    \centering
    \includegraphics[scale=0.75]{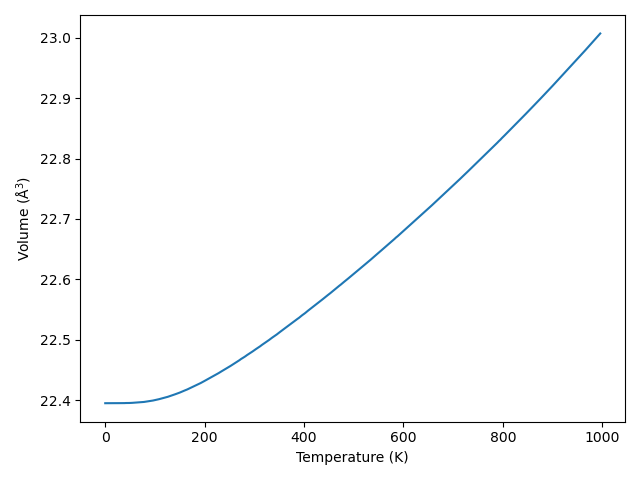}
    \caption{Change in volume of Fe with temperature as calculated by \elagenteS{} with Phonopy. \elagenteS{} generated this plot.}
\end{figure*}

\begin{figure*}[htbp]
    \centering
    \includegraphics[scale=0.75]{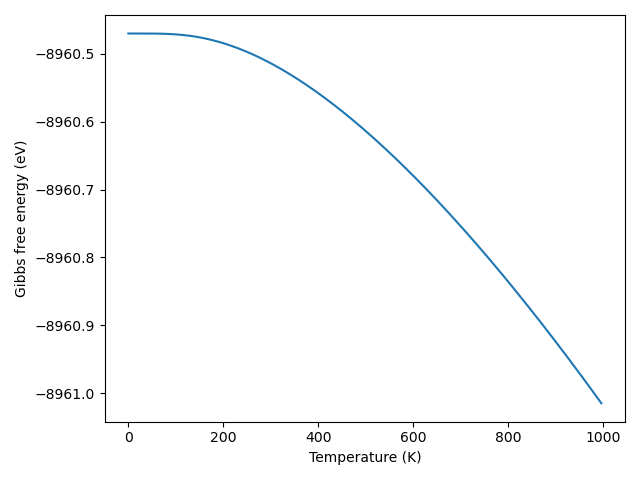}
    \caption{Change in Gibbs free energy of Fe with temperature as calculated by \elagenteS{} with Phonopy. \elagenteS{} generated this plot.}
\end{figure*}

\begin{figure*}[htbp]
    \centering
    \includegraphics[scale=0.75]{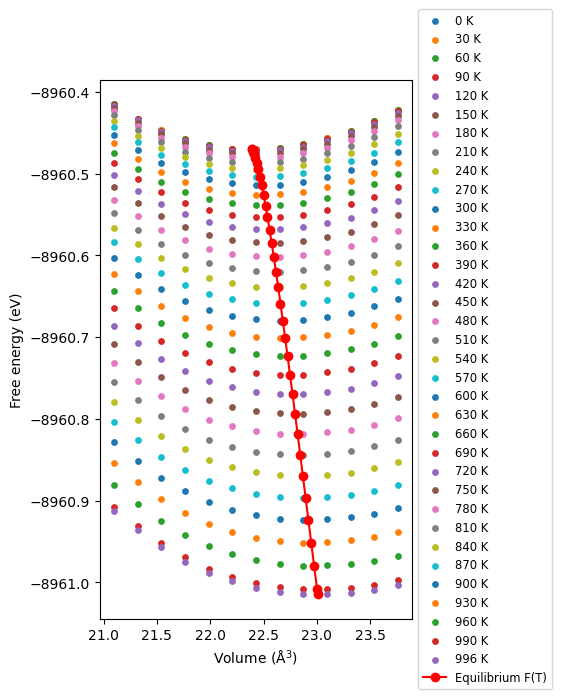}
    \caption{Change in Helmholtz free energy of Fe with temperature and volume as calculated by \elagenteS{} with Phonopy. \elagenteS{} generated this plot.}
\end{figure*}

\begin{figure*}[htbp]
    \centering
    \includegraphics[scale=0.75]{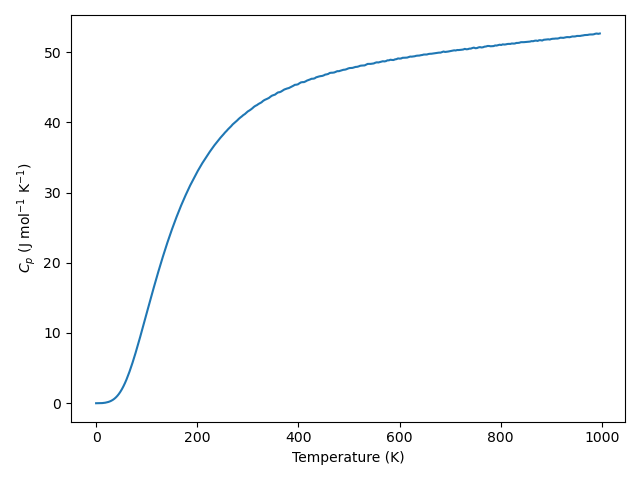}
    \caption{Change in heat capacity (C$_p$) of Fe with temperature and volume as calculated by \elagenteS{} with Phonopy. \elagenteS{} generated this plot.}
\end{figure*}

\begin{figure*}[htbp]
    \centering
    \includegraphics[scale=0.75]{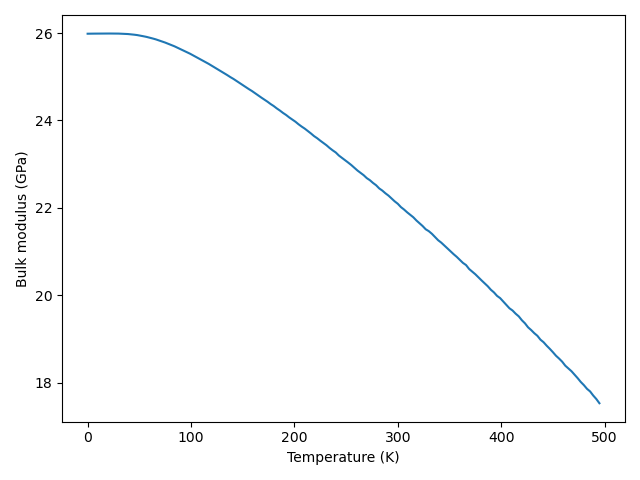}
    \caption{Change in bulk modulus of NaCl with temperature as calculated by \elagenteS{} with Phonopy. \elagenteS{} generated this plot.}
\end{figure*}

\begin{figure*}[htbp]
    \centering
    \includegraphics[scale=0.75]{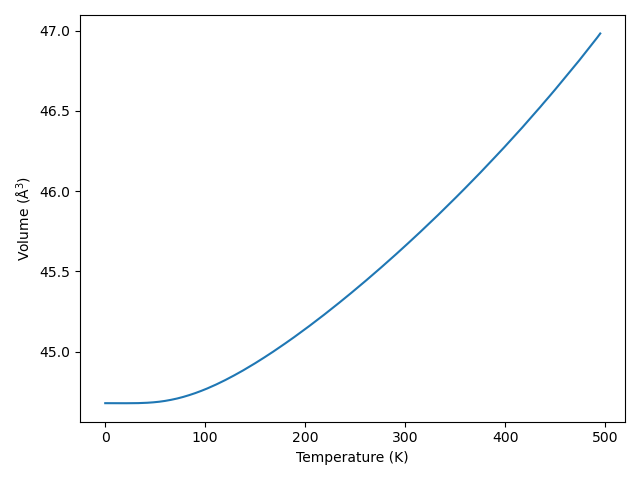}
    \caption{Change in volume of NaCl with temperature as calculated by \elagenteS{} with Phonopy. \elagenteS{} generated this plot.}
\end{figure*}

\begin{figure*}[htbp]
    \centering
    \includegraphics[scale=0.75]{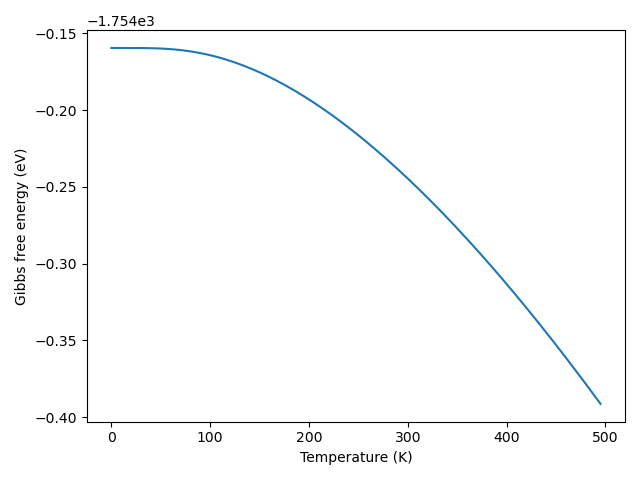}
    \caption{Change in Gibbs free energy of NaCl with temperature as calculated by \elagenteS{} with Phonopy. \elagenteS{} generated this plot.}
\end{figure*}

\begin{figure*}[htbp]
    \centering
    \includegraphics[scale=0.75]{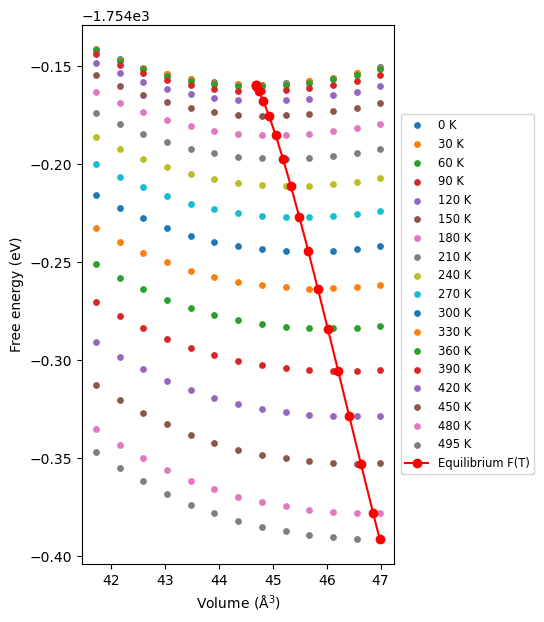}
    \caption{Change in Helmholtz free energy of NaCl with temperature and volume as calculated by \elagenteS{} with Phonopy. \elagenteS{} generated this plot.}
\end{figure*}

\begin{figure*}[htbp]
    \centering
    \includegraphics[scale=0.75]{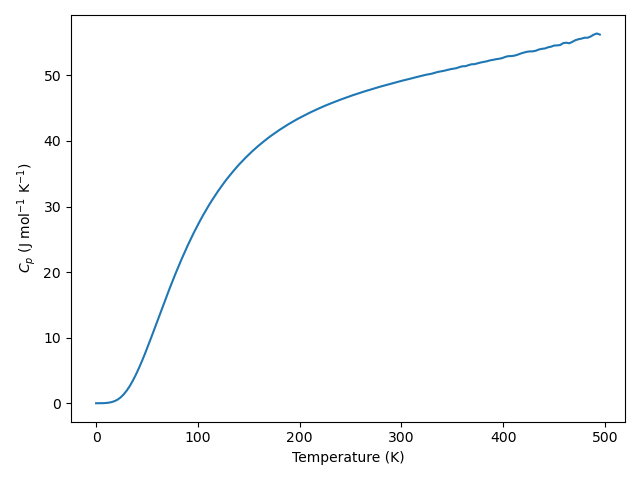}
    \caption{Change in heat capacity (C$_p$) of NaCl with temperature and volume as calculated by \elagenteS{} with Phonopy. \elagenteS{} generated this plot.}
\end{figure*}

\FloatBarrier

\subsection{Figures for MOF case study}

\begin{figure*}[htbp]
    \centering
    \includegraphics[width=0.98\textwidth]{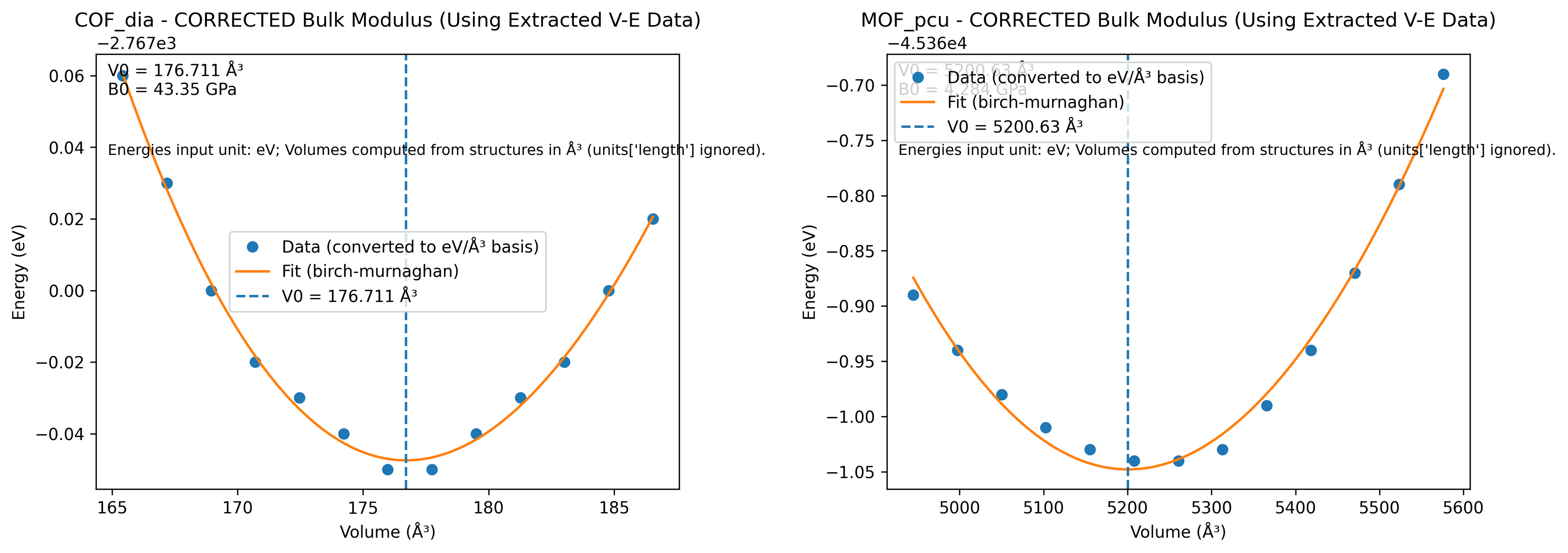}
    \caption{For the two reticular frameworks, \elagenteS{} generated for each, from the data in the tables above, an energy-volume curve along with a Birch-Murnaghan fit.}
    \label{mof_energy_volume}
\end{figure*}

\begin{figure*}[htbp]
    \centering
    \includegraphics[scale=0.75]{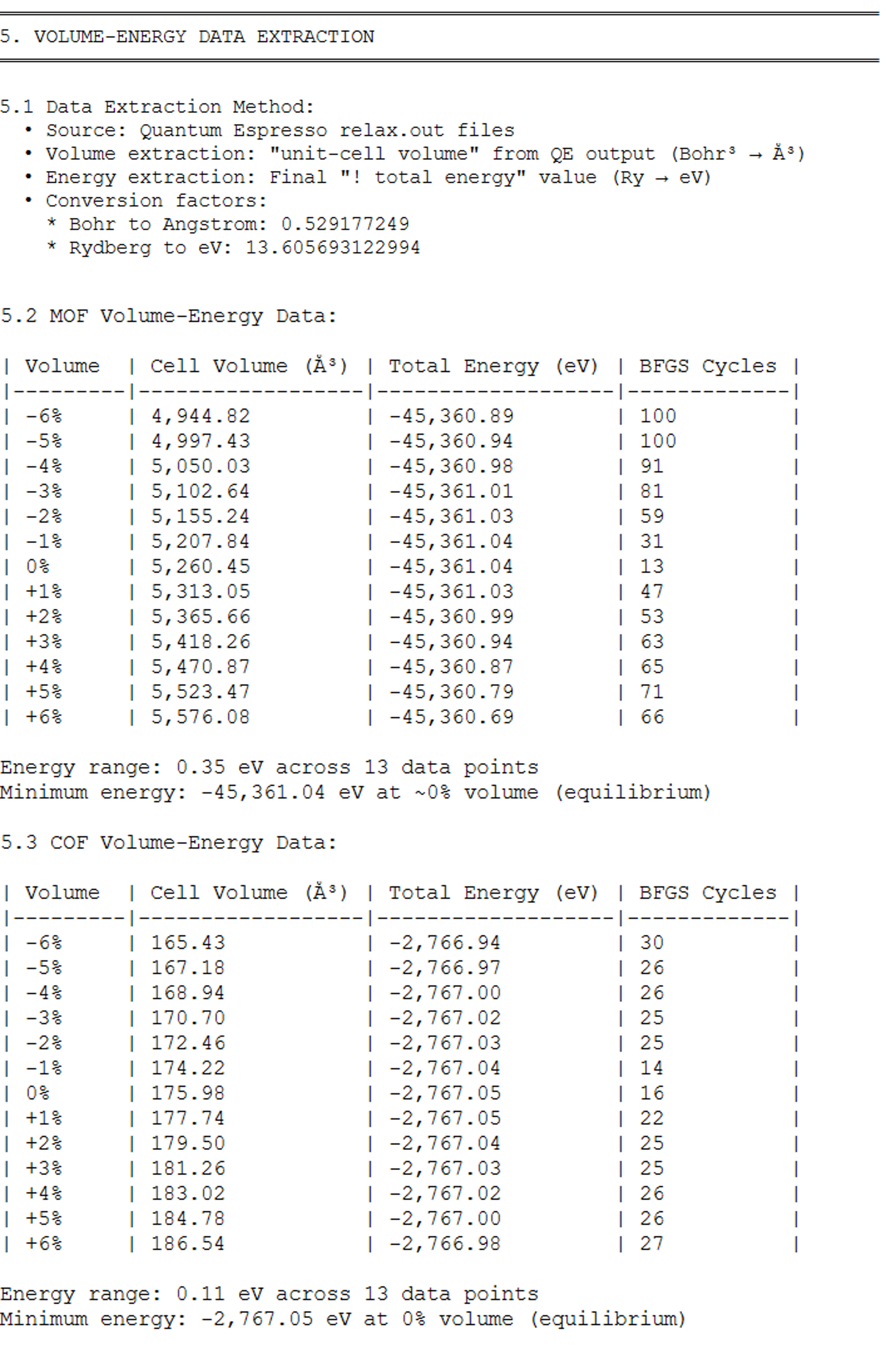}
    \caption{\elagenteS{} extracted the energy, volume, and number of BFGS steps for each structure in the MOF/COF case study and organized them into two separate tables in the summary report.}
    \label{mof_section_5}
\end{figure*}

\endgroup

%% file: includes/lit_review_table.tex

\begin{longtable}{
    @{\extracolsep{\fill}}
    >{\RaggedRight\hspace{0pt}}p{0.12\textwidth}  
    >{\RaggedRight\hspace{0pt}}p{0.25\textwidth}  
    >{\RaggedRight\hspace{0pt}}p{0.18\textwidth}  
    >{\RaggedRight\hspace{0pt}}p{0.30\textwidth}  
    >{\centering\arraybackslash}p{0.05\textwidth} 
}

\caption{Overview of agentic systems, their features, external tools, and benchmarks.} \label{tab:agents} \\
\toprule
\thead{\textbf{System}} & 
\thead{\textbf{Main} \\ \textbf{Features}} & 
\thead{\textbf{Primary External} \\ \textbf{Code(s)}} & 
\thead{\textbf{Computational} \\ \textbf{Benchmarks}} & 
\thead{\textbf{Ref.}} \\
\midrule
\endfirsthead

\multicolumn{5}{c}%
{{\bfseries \tablename\ \thetable{} -- continued from previous page}} \\
\toprule
\textbf{System} & \textbf{Main Features} & \textbf{Primary External Code(s)} & \textbf{Computational Benchmarks} & \textbf{Ref.} \\
\midrule
\endhead

\midrule
\multicolumn{5}{r}{{Continued on next page}} \\
\bottomrule
\endfoot

\bottomrule
\endlastfoot

MOFGen & Platform for MOF discovery & MACE, XTB, VASP & Experimental validation of 5 ``AI-dreamt'' MOFs & \cite{mofgen2025} \\
\addlinespace

MAPPS & Crystal structure generation and prediction & ASE, Pymatgen, CHGNet & Matching properties of generated structures against existing databases & \cite{MAPPS2025} \\
\addlinespace

DREAMS & Platform for doing solid-state DFT calculations & Quantum Espresso, ASE & Lattice constants against Sol27LC; CO adsorption against literature & \cite{dreams2025} \\
\addlinespace

SciLink & Connects experiments and simulation for novelty assessment & VASP, ASE & n/a & \cite{SciLink2025} \\
\addlinespace

ChatMOF & MOF generation and analysis with forward/inverse design & PORMAKE, MOFTransformer, ASE & Set of sample questions for workflow logic correctness & \cite{chatmof2024} \\
\addlinespace

CatMaster & Establishing and running workflows for heterogeneous catalysis & VASP, MACE, Pymatgen & Surface energies of Fe low index facets against literature & \cite{catmaster2026} \\
\addlinespace

VASPilot & Framework for doing VASP calculations & VASP, Pymatgen & Lattice constants of 2H-MoS\textsubscript{2} against experimental values & \cite{vaspilot2025} \\
\addlinespace

Adsorb-Agent & Identifying stable adsorption configurations of molecules & EquiformerV2, Pymatgen & Adsorption energies (vs enumeration algorithms); consistency across trials & \cite{adsorbagent2025} \\
\addlinespace

AGAPI-Agents & Platform for prediction, search, and visualization of crystals & JARVIS-Tools, ASE, ALIGNN, SlaKoNet & Bulk modulus, band gap, Tc, etc. against JARVIS-Leaderboard & \cite{AGAPIAgents2025} \\
\addlinespace

GENIUS & Knowledge graph-based framework for Quantum Espresso & Quantum Espresso & Completion rate of 295 prompts generated by human experts & \cite{genius2025} \\
\addlinespace

Crystalyse & Platform for generating, optimizing, and predicting inorganic crystals & SMACT, Chemeleon, MACE, ASE, Pymatgen & Accuracy of predicted labels against in-house dataset & \cite{crystalyse2025} \\
\addlinespace

AURA & Agentic system that interfaces with nanoHUB & Packages accessible via nanoHUB & n/a & \cite{AURA2025} \\
\addlinespace

LLMatDesign & Platform for predicting materials with desired property via modifications & ASE, VASP, in-house MLFF & LLM-guided vs. random modifications &\cite{LLMatDesign2024} \\
\addlinespace

MatSciAgent & Framework for generating structures and continuum/molecular dynamics & ASE, CrystalLLM & n/a & \cite{matsciagent2025} \\
\addlinespace

Vriza et al. & Platform for doing molecular dynamics simulations & LAMMPS, Phonopy, ASE, Atomsk & Lattice constants, elastic constants, melting temps vs literature & \cite{VrizaEtAl2025} \\
\addlinespace

AtomAgents & Framework for atomistic simulations for alloy design & LAMMPS & n/a & \cite{atomagents2024} \\
\addlinespace

Masagent & A self-contained python package for doing VASP calculations & VASP, various MLIPs (SevenNet, CHGNet, etc.) & Energy/atom for various materials, computational scaling & \cite{Masagent2025} \\
\addlinespace

QUASAR & A containerized platform for performing solid-state DFT and molecular dynamics simulations & Quantum ESPRESSO, MACE, LAMMPS, RASPA3, ASE, Pymatgen & K-point convergence and a variety of properties including band gap and melting point & \cite{yang_quasar_2026} \\
\addlinespace

K-Dense & Claude equipped with scientific skills for R\&D & n/a & n/a & \cite{KDensePaper2025, KDenseCompany2025} \\
\addlinespace

Paramus & General materials science R\&D workflows with inverse design & ORCA, LAMMPS, AIMNet2, MatterGen & n/a & \cite{ParamusCompany2025} \\
\addlinespace

El Agente S\'olido & Platform for modelling solid-state materials & Quantum Espresso, ASE, Pymatgen, ATAT, MACE, UMA & See Table 1 of main text & This Work \\

\end{longtable}